\documentclass{egpubl}
\usepackage{pg2023}

\SpecialIssuePreprint    % uncomment for submission to , special issue
\usepackage[T1]{fontenc}
\usepackage{dfadobe}  

\usepackage{cite}  % comment out for biblatex with backend=biber
\BibtexOrBiblatex
\electronicVersion
\PrintedOrElectronic
\ifpdf \usepackage[pdftex]{graphicx} \pdfcompresslevel=9
\else \usepackage[dvips]{graphicx} \fi

\usepackage{egweblnk} 
\usepackage{amsmath}
\usepackage{amssymb}

\usepackage{booktabs, multirow} % for borders and merged ranges
\usepackage{soul}% for underlines
\usepackage{changepage,threeparttable} % for wide tables
\usepackage[table]{xcolor}

\usepackage{import}
\usepackage{xifthen}
\usepackage{pdfpages}
\usepackage{transparent}

\title[Neural Impostor: Editing Neural Radiance Fields with Explicit Shape Manipulation]%
      {Neural Impostor: \\ Editing Neural Radiance Fields with Explicit Shape Manipulation}

\author[R. Liu, J. Xiang, B. Zhao, R. Zhang, J. Yu \& C. Zheng]
{\parbox{\textwidth}{\centering Ruiyang Liu\footnotemark[1]$^{1,2}$\orcid{0000-0003-0075-6230}, Jinxu Xiang\footnotemark[1]$^{1}$\orcid{0009-0003-6230-6048}, Bowen Zhao$^{1}$\orcid{0000-0001-6959-6963}, Ran Zhang$^{*1}$\orcid{0000-0002-3808-281X}, Jingyi Yu$^{2}$ and Changxi Zheng$^{1}\orcid{0000-0001-9228-1038}$
        }
        \\
{\parbox{\textwidth}{\centering $^1$Tencent Pixel Lab, $^2$ShanghaiTech University, \footnotemark[1]Equally contributed co-first authors
       }
}
}

\newcommand{\figref}[1]{Figure~\ref{fig:#1}}
\newcommand{\tabref}[1]{Table~\ref{tab:#1}}
\newcommand{\eqnref}[1]{Equation~\eqref{eq:#1}}
\newcommand{\secref}[1]{Section~\ref{sec:#1}}

\makeatletter
\DeclareRobustCommand\onedot{\futurelet\@let@token\@onedot}
\def\@onedot{\ifx\@let@token.\else.\null\fi\xspace}

\makeatother

\renewcommand{\d}[1]{\ensuremath{\operatorname{d}\!{#1}}}

\begin{document}

% uncomment for using teaser
\teaser{
    \def\svgwidth{2\columnwidth}
    %% Creator: Inkscape 1.2.2 (732a01da63, 2022-12-09), www.inkscape.org
%% Accompanies image file 'teaser.pdf' (pdf, eps, ps)
\begingroup%
  \makeatletter%
  \providecommand\color[2][]{%
    \errmessage{(Inkscape) Color is used for the text in Inkscape, but the package 'color.sty' is not loaded}%
    \renewcommand\color[2][]{}%
  }%
  \providecommand\transparent[1]{%
    \errmessage{(Inkscape) Transparency is used (non-zero) for the text in Inkscape, but the package 'transparent.sty' is not loaded}%
    \renewcommand\transparent[1]{}%
  }%
  \providecommand\rotatebox[2]{#2}%
  \newcommand*\fsize{\dimexpr\f@size pt\relax}%
  \newcommand*\lineheight[1]{\fontsize{\fsize}{#1\fsize}\selectfont}%
  \ifx\svgwidth\undefined%
    \setlength{\unitlength}{1843bp}%
    \ifx\svgscale\undefined%
      \relax%
    \else%
      \setlength{\unitlength}{\unitlength * \real{\svgscale}}%
    \fi%
  \else%
    \setlength{\unitlength}{\svgwidth}%
  \fi%
  \global\let\svgwidth\undefined%
  \global\let\svgscale\undefined%
  \makeatother%
  \begin{picture}(1,0.43407488)%
    \lineheight{1}%
    \setlength\tabcolsep{0pt}%
    \put(0,0){\includegraphics[width=\unitlength,page=1]{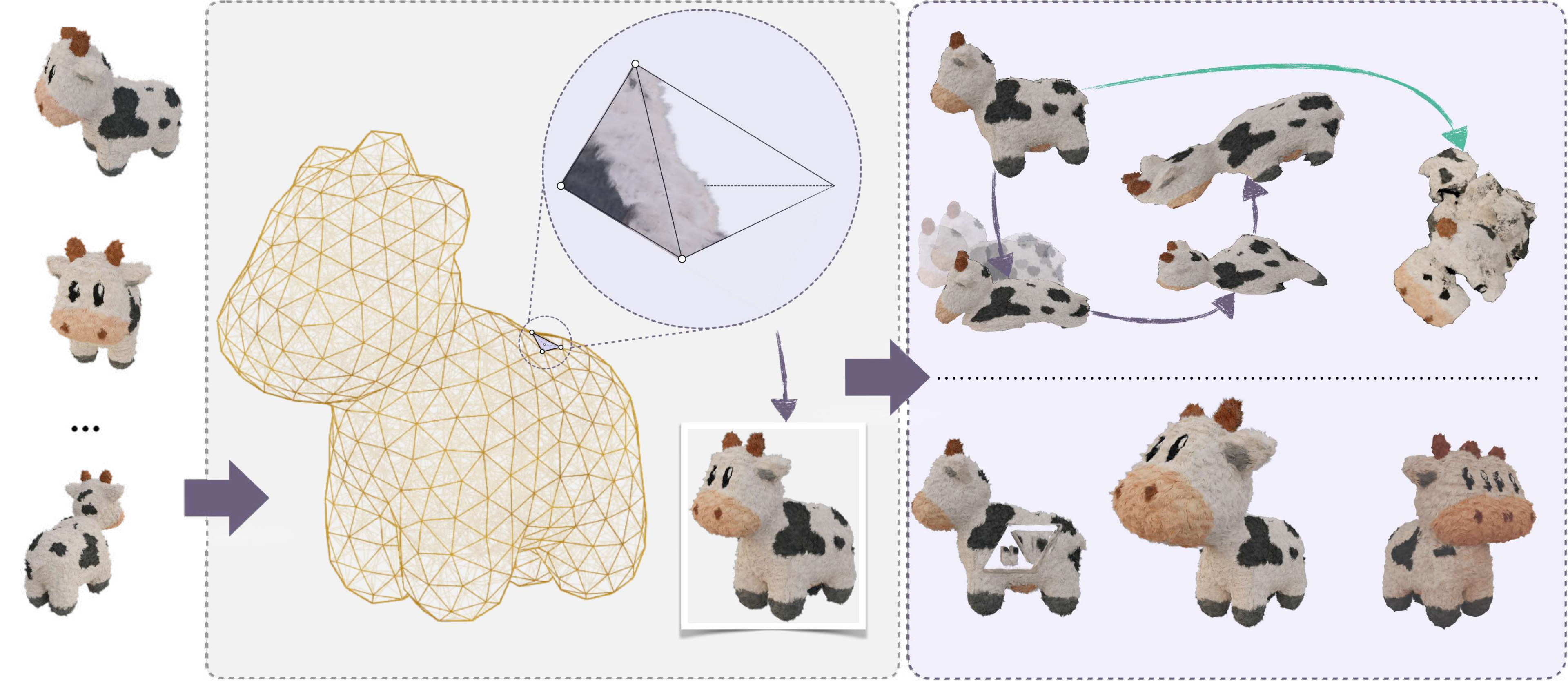}}%
    \put(0.06739618,0.02107163){\color[rgb]{0,0,0}\makebox(0,0)[t]{\lineheight{1.25}\smash{\begin{tabular}[t]{c}Multi-view\\Images\end{tabular}}}}%
    \put(0.14784765,0.40966471){\color[rgb]{0,0,0}\makebox(0,0)[lt]{\lineheight{1.25}\smash{\begin{tabular}[t]{l}\textbf{Neural Impostor}\end{tabular}}}}%
    \put(0.28737524,0.01293273){\color[rgb]{0,0,0}\makebox(0,0)[t]{\lineheight{1.25}\smash{\begin{tabular}[t]{c}Explicit Tetrahedral Proxy\end{tabular}}}}%
    \put(0.45026128,0.25718102){\color[rgb]{0,0,0}\makebox(0,0)[t]{\lineheight{1.25}\smash{\begin{tabular}[t]{c}Barycentric\\Encoding\end{tabular}}}}%
    \put(0.71533468,0.20309533){\color[rgb]{0,0,0}\makebox(0,0)[t]{\lineheight{1.25}\smash{\begin{tabular}[t]{c}Soft Body\end{tabular}}}}%
    \put(0.65245118,0.01293273){\color[rgb]{0,0,0}\makebox(0,0)[t]{\lineheight{1.25}\smash{\begin{tabular}[t]{c}Boolean Operation\end{tabular}}}}%
    \put(0.78631469,0.01189682){\color[rgb]{0,0,0}\makebox(0,0)[t]{\lineheight{1.25}\smash{\begin{tabular}[t]{c}Deformation\end{tabular}}}}%
    \put(0.92505499,0.01303871){\color[rgb]{0,0,0}\makebox(0,0)[t]{\lineheight{1.25}\smash{\begin{tabular}[t]{c}Composition\end{tabular}}}}%
    \put(0.92737446,0.20228991){\color[rgb]{0,0,0}\makebox(0,0)[t]{\lineheight{1.25}\smash{\begin{tabular}[t]{c}Fracture\end{tabular}}}}%
    \put(0.49267065,0.01296187){\color[rgb]{0,0,0}\makebox(0,0)[t]{\lineheight{1.25}\smash{\begin{tabular}[t]{c}Rendered Image\end{tabular}}}}%
    \put(0.9879059,0.40966471){\color[rgb]{0,0,0}\makebox(0,0)[rt]{\lineheight{1.25}\smash{\begin{tabular}[t]{r}\textbf{Physical Simulation}\end{tabular}}}}%
  \end{picture}%
\endgroup%

 \centering
  \caption{The creation and application of Neural Impostor. \textit{Neural Impostor} takes multi-view captured images as input and build a coarse geometry proxy with local radiance fields encoded within the barycentric coordinate. With the help of the coarse geometry proxy, Neural Impostor empowers novice users to conduct vivid physical simulations and perform versatile editing operations on radiance fields, including Boolean operation, local deformation and composition.}
\label{fig:teaser}
}

\maketitle
%-------------------------------------------------------------------------
\begin{abstract}
% Neural Radiance Fields (NeRF) have shown remarkable promise in generating highly realistic and expressive 3D scenes.
% While NeRF has made significant strides in accurately modeling a scene's geometry and appearance, it remains challenging to edit, especially when it comes to changing the scene's geometry.
% This limitation hinders the broader adoption of NeRF in a variety of applications.
% In light of this, we present Neural Impostor, a novel editable NeRF-like representation that utilizes a hybrid formulation, which includes an explicit tetrahedral mesh and a multi-grid implicit field for every tetrahedron in the tetrahedral mesh.
% The proposed framework leverages the existing editing operations of  explicit proxies, as well as the convenience of reconstructing complex volumetric appearance with implicit fields, and provides a pragmatic tool for deforming, compositing and generating neural radiance creations.
% This framework relies on a multigrid barycentric representation of an explicit cage-like proxy.
% This representation enables not only continuous editing operations like deformation,
% but also topological operations, such as remeshing the proxies.
% The capable editing operations then form a comprehensive editing pipeline of neural implicit fields.
% We demonstrate the robustness and versatility of our system with various examples and experiments,
% including editing synthetic objects and real captured data.
% Furthermore, we show the authoring process of a hybrid synthetic-captured object with various editing operations.

Neural Radiance Fields (NeRF) have significantly advanced the generation of highly realistic and expressive 3D scenes.
However, the task of editing NeRF, particularly in terms of geometry modification, poses a significant challenge.
This issue has obstructed NeRF's wider adoption across various applications. To tackle the problem of efficiently editing neural implicit fields, we introduce \textbf{Neural Impostor},
a hybrid representation incorporating an explicit tetrahedral mesh alongside a multigrid implicit field designated for each tetrahedron within the explicit mesh. 
Our framework bridges
% editing operations on the explicit proxy and the implicit fields
the explicit shape manipulation and the geometric editing of implicit fields
by utilizing multigrid barycentric coordinate encoding, thus offering a pragmatic solution to deform, composite, and generate neural implicit fields while maintaining a complex volumetric appearance.
Furthermore, we propose a comprehensive pipeline for editing neural implicit fields based on a set of explicit geometric editing operations. We show the robustness and adaptability of our system
through diverse examples and experiments, including the editing of both synthetic objects and real captured data.
Finally, we demonstrate the authoring process of a hybrid synthetic-captured object utilizing a variety of editing operations, underlining the transformative potential of \textbf{Neural Impostor} in the field of 3D content creation and manipulation.

%-------------------------------------------------------------------------
%  ACM CCS 1998
%  (see https://www.acm.org/publications/computing-classification-system/1998)
% \begin{classification} % according to https://www.acm.org/publications/computing-classification-system/1998
% \CCScat{Computer Graphics}{I.3.3}{Picture/Image Generation}{Line and curve generation}
% \end{classification}
%-------------------------------------------------------------------------
%  ACM CCS 2012
%  (see https://www.acm.org/publications/class-2012)
%The tool at \url{http://dl.acm.org/ccs.cfm} can be used to generate
% CCS codes.
%Example:


\end{abstract}

\section{Introduction}

% \cmt{NeRF becomes increasingly more powerful and expressive}

% \cmt{NeRF remains difficult to edit and animate, e.g., edit geometry. Prevent it from a wider range of use}

% \cmt{We focus on the edit of geometry. Our proposed idea: tet mesh + NeRF field}

% \cmt{Editing operation causes local changes -> Tet mesh is more suitable for local changes}

The Neural Radiance Field (NeRF) serves as a groundbreaking neural implicit representation that revolutionizes novel view synthesis processes. It allows users to rebuild a scene directly from multi-view photographs using a standard camera through an end-to-end training process. This significant advancement has streamlined complex 3D reconstruction processes, signifying a considerable paradigm shift in the field. However, the transformation of 3D scene data into a higher-dimensional feature space implicitly stored within neural networks presents notable challenges, including incompatibility with most existing modeling and editing software. As a result, editing neural radiance fields, especially in terms of geometry modification, remains a considerable challenge.

Recent research in editing neural radiance fields have been split primarily into two categories: implicit and explicit mapping.
Implicit methods, though widely adopted, are often limited due to their dependence on high-dimensional feature space interpolation, restricted to accommodate only simple edits.
On the other hand, explicit methods that employ dedicated proxies for direct manipulation of the radiance field provide enhanced editing capabilities, albeit with inherent challenges such as modeling limitations and increased computational complexity.
For example, while \textit{triangle meshes based methods}~\cite{Chen2022MobileNeRFET,neumesh2022,neumanifold2023} favored for their compatibility with existing 3D modeling tools and high-speed rendering capabilities, they struggle with modeling volumetric appearances like furs and hairs. The employment of \textit{point clouds}~\cite{pointnerf2022,Chen2023NeuralEditorEN} enables volumetric modeling but lacks geometry constraints, complicating contiguous authoring.
The utilization of \textit{volumetric cages}~\cite{voltemorph2022,nerf-editing2022,cage2022,Peng2022CageNeRF,Jambon2023NeRFshop}, on the other hand, offers topology constraints for stable authoring and benefits in physical simulation. Despite these advantages, their rendering speed is often hampered by complex point-in-tetrahedron queries.

In response to these challenges, we introduce Neural Impostor, a hybrid model that addresses the primary considerations necessary for creating a fully editable neural radiance field that compatible with common geometric editing pipeline.
It is also capable of high fidelity reconstruction and optimized for real-time rendering.

The Neural Impostor segments the modeling space of NeRF with explicit tetrahedral meshes, transitioning from Cartesian space encoding to local barycentric encoding within the tetrahedral proxy.
Each tetrahedron is assigned a distinct Multi-grid Neural Radiance Field for detailed local depictions, enhancing scalability in the modeling and rendering process.
Our barycentric encoding scheme maintains visual uniformity during physical simulations driven by vertex transformations due to its harmonic nature around $n+1$ control points.
Furthermore, Neural Impostor's compatibility with existing 3D animation software, such as Houdini, offers distinct advantages in animations and games.
Drawing inspiration from space partitioning concepts, our model facilitates high-quality, real-time rendering and is crucial for modeling, rendering, and simulating complex volumetric objects like plush toys.

In essence, the Neural Impostor framework brings together the advantages of explicit mesh editing with the volumetric appearance of implicit fields, thereby offering a powerful and versatile tool for manipulating Neural Radiance Fields. We highlight our key contributions as follows:
\begin{itemize}
  \item The design and implementation of \textit{Neural Impostor}, a novel hybrid representation that bolsters the editing capabilities of Neural Radiance Fields by adeptly leveraging the benefits of both explicit meshes and implicit fields.
  \item The deployment of a \textit{Robust Hash Encoding} method combined with \textit{Sophisticated Spatial Sampling} techniques. By utilizing barycentric coordinates encoding of impostor geometric elements, we ensure high-quality reconstruction and rendering during animations.
  \item The introduction of a diverse suite of \textit{Geometric Operations} underpinned by the \textit{Neural Impostor}. This approach enables the seamless transfer of local geometry editing operations, such as shape morphing, remeshing and boolean operations, from contemporary 3D modeling software into Neural Radiance Fields through local retraining.
  \item The creation of an efficient, hardware-accelerated rendering algorithm, which streamlines the traditionally laborious barycentric coordinate mapping process. This algorithm fulfills real-time requirements, achieving an impressive performance of approximately 30 FPS for $1080\times 1080$ rendering.
\end{itemize}

In the forthcoming sections, we will first delve into the modeling of \textit{Neural Impostor} (\secref{modeling})).
Alongside this, we detail a set of editing operations for Neural Impostor that accommodate edits at the animation, geometric, and appearance levels (\secref{operations}).
We then demonstrate the modeling and editing capabilities of \textit{Neural Impostor} through systematic testing on the \textit{nerf-synthetic} dataset, as well as on a few custom-built \textit{plush toy} models with complex volumetric appearances(\secref{expr}). Additionally, we will present operations such as shape morphing, soft body deformation, fracture, and composition made possible by \textit{Neural Impostor}, thereby effectively underlining the utility of \textit{Neural Impostor} in real-world content creation workflows.
This will be exemplified through a practical demonstration of constructing a snowman toy by editing multiple pre-trained Neural Impostors (\secref{application}).

\section{Related Work}
Neural Impostor builds upon the foundation of Neural Radiance Fields (NeRF)~\cite{Mildenhall2020NeRFRS}, which presents a method for representing a 3D scene using a scalar density field $\sigma$ and a view-dependent color field $\mathbf{c}$.
These fields enable NeRF to generate high-quality renderings from novel viewpoints using volume rendering techniques like \cite{Max1995OpticalMF}.
The core idea of NeRF is to employ an implicit representation due to the inherent complexity of capturing both 3D geometry and appearance.
This representation involves encoding the spatial information, which typically consists of positional parameter $\mathbf{x}$ and directional parameter $\mathbf{d}$, into a learnable function $f_{\mathbf{\theta}}:(\mathbf{x,d}) \mapsto (\mathbf{c},\sigma)$ with adjustable parameters $\mathbf{\theta}$.
With a space sampler $\mathcal{S}$, the novel view appearance can be rendered by integrating ($\mathbf{c},\sigma$) along the sampling ray $\mathbf{r}$:

\begin{equation}
    \mathcal{C}(\mathbf{r}) = \sum_{i=1}^N T_i(1-\exp (\sigma_i,\delta_i))c_i,\textit{ where }T_i=\exp\bigl( -\sum_{j=1}^i\sigma_j\delta_j\bigr)
\end{equation}

where $\delta_i$ denotes the sampling intervals along the ray $\mathbf{r}$ and the transmittance $T_i$ is the accumulated density along the ray.
With multi-view images as supervision, the learnable function $f_{\mathbf{\theta}}$ can be trained by minimizing the reconstruction loss between the rendered image and the ground truth image.
% Neural Impostor expands upon Neural Radiance Fields(NeRF)~\cite{Mildenhall2020NeRFRS}. 
% NeRF represents a 3D scene by a scalar density field and a view dependent radiance field.
% With volume rendering techniques such as \cite{Max1995OpticalMF}, NeRF can render high-quality appearances from novel viewpoints.
% Due to the complexity of representing 3D geometry and appearance, NeRF uses an implicit representation.
% By encoding the space $\mathcal{G}$,
% which is usually the positional parameters and directional parameters, into a function $f_{\theta}$ that possesses learnable parameters $\theta$.

Recent surveys on NeRF \cite{STARNR2022} and \cite{STARNR2022} presented a comprehensive review of every process in computing NeRF.
This paper primarily concentrates on the geometric editing of NeRF among all the stages involved in its construction and manipulation.
Rather than providing an exhaustive overview of all existing NeRF research, we specifically focus on reviewing the approach that directly pertains to our topic.
\subsection{Spatial Encoding in Neural Radiance fields}

When it comes to geometric editing of 3D content, the core of the problem is to encode and manipulate the spatial information of the scene.
The spatial encoding process of NeRF is to map the low-dimensional smooth inputs such as position and direction to a high-dimensional feature space that can be trained to encode high-frequency details.
Initially, researchers use a fully connected MLP with positional encoding~\cite{Mildenhall2020NeRFRS} to encode the spatial information of the entire scene, which synthesizes realistic novel views with fine details, but also suffers from high computational cost when querying the large neural network for every sample.
To enhance the training efficiency, hybrid spatial representations utilizes the sparsity of spatial data, factorize the scene into explicitly stored features and implicit neural functions.
Depending on the underlying structure, we categorize spatial encoding methods into grid-based and mesh-based structures.
Grid-based encoding schemes, here we refer to pre-constructed regular spatial structures, such as sparse voxel grids~\cite{NSVF2020,SNeRG2021}, octree grids~\cite{Yu2021PlenOctreesFR,Yu2021PlenoxelsRF}, triplanar structures~\cite{EG3D2021}, codebooks~\cite{VQAD2022}, tensorial decomposition~\cite{Chen2022TensoRFTR} and multi-grid hash encoding~\cite{Mller2022InstantNG}.
The grid-based approaches provide efficient querying of features stored in the grid, which greatly improves the training and rendering speed.
Mesh-based encoding, on the other hand, uses a mesh-like structure with more flexible topologies, such as point cloud~\cite{pointnerf2022}(point clouds can be seen as unconnected mesh vertices), triangle soup~\cite{Chen2022MobileNeRFET}, manifold mesh~\cite{neumesh2022,neumanifold2023}, duplex mesh~\cite{duplexnerf2023} and tetrahedral mesh~\cite{gao2020deftet}.
Mesh-based encoding provides additional flexibility in manipulating implicit fields; however, its irregular structure necessitates a robust adaptive mesh generator. Furthermore, when storing all the features on explicit meshes, mesh-based encoding often lacks volumetric appearances.

In this paper, we aim to achieve both the efficiency of grid-based encoding and the flexibility of mesh-based encoding, therefore enables the geometric editing of NeRF while maintaining the high-quality volumetric rendering.
In the proposed method, we utilize a tetrahedral based proxy, and bounds the explicit mesh and implicit field with a generalized barycentric encoding method.

\subsection{Proxy-based manipulation of 3D models}
Using explicit proxies to manipulate 3D models is a common practice in computer graphics.
Proxies usually retain a lower dimensional representation of the original model, which serve as a model reduction process and can be used to accelerate the rendering process or provide a more intuitive interface for users to manipulate the model.
For example, in geometric modeling and shape manipulation, proxies reduces the complexity of original models,
therefore enables structure preserving shape manipulation~\cite{Zheng2011CompController}, shape collection synthesis~\cite{Xu2012Evolution} and image-based geometric modeling~\cite{Xu2011PhotoModeling}.
Bounding volumes, which are one of the simplest form of proxies, are widely used in collision detection~\cite{Gottschalk1996OBBTreeAH,Klosowski1998EfficientCD}, ray tracing~\cite{meister2021BVHsurvey} and shape approximation~\cite{Calderon:2017:BPS}.
Proxy-based methods are also used in manipulating NeRF, one of the commonly used proxies is the tetrahedral mesh, which is used to approximate the geometry of the scene.
By mapping the samples in the deformed tetrahedral mesh to the canonical space~\cite{voltemorph2022,cage2022,nerf-editing2022,Peng2022CageNeRF}, one can manipulate the neural implicit field with geometric operations on the explicit tetrahedral mesh.
In this paper, we follow the general idea of using tetrahedral mesh as proxy, but instead of bending sampling rays to the canonical space, we propose a generalized barycentric encoding method efficiently store the implicit field in the tetrahedral mesh, which allows more flexible operations,
please see Sec.~\ref{sec:operations}.
Worth to mention that there is a special type of proxy called impostor\cite{jeschke05ARI}, which originally is a 2D image that approximates the appearance of a 3D model from a specific viewpoint by interpolating pre-rendered multi-view appearances.
The concept of impostor is later extended to 3D models, such as \cite{jeschke2002textured,yee2008crowd}, and is widely used in real-time rendering of large-scale scenes.
In this paper, we introduce the term, "Neural Impostor," which is derived from the concept of Impostor.
Our proposed method entails the creation of an editable 3D proxy that approximates the volumetric appearance of a 3D model from arbitrary viewpoints, just like traditional impostor did in real-time rendering.

\subsection{Editing Neural Radiance Fields}
Recently, Editing Neural Radiance Fields has seen significant progress in terms of reconstructing and modifying various aspects of appearance, such as relighting, controlling shapes, and altering colors or palettes of objects\cite{Liu2021EditingCR,Kuang2022PaletteNeRFPA}. Some techniques~\cite{nerf-editing2022,voltemorph2022,Jambon2023NeRFshop} have enabled modification of scene parts and their respective locations, thus providing more control over the rendered scene. However, editing the dynamics of moving objects, especially with topological changes, has been a major challenge. These changes can lead to motion discontinuities, causing noticeable artifacts if not properly modeled\cite{nerf-editing2022}. \cite{Kania2021CoNeRFCN} has tried to mitigate these challenges using manual supervision, but its capabilities are limited to one-dimensional editing per scene part, necessitating user annotations for supervision. In contrast, EditableNeRF\cite{Zheng2022EditableNeRFET} uses an image sequence from a single camera to train a network, modeling topologically varying dynamics via surface key points. Users can edit the scene by manipulating these key points, enabling more intuitive multi-dimensional editing (up to 3D). NeuralEditor\cite{Chen2023NeuralEditorEN} is another novel approach that leverages the explicit point cloud representation underlying NeRFs for shape editing tasks. It employs a unique rendering scheme based on deterministic integration within K-D tree-guided density-adaptive voxels, which results in high-quality rendering and precise point clouds. Despite these advancements, editing capabilities within a computed NeRF scene remain relatively limited, especially when compared to traditional CGI workflows.
Recent progress in pre-trained large-scale models has enabled rapid progress in a brand new instruction or prompt based creation and interaction with digital contents. For instance, Instruct-NeRF2NeRF\cite{Haque2023InstructNeRF2NeRFE3} leverages an image-conditioned diffusion model to iteratively edit input images while simultaneously optimizing the underlying radiance field of the 3D scene, resulting in a realistic 3D scene that adheres to the provided editing instructions. \cite{Mendiratta2023AvatarStudioTE} offering genuine, personalized, temporal consistent 3D avatar edit by combining text-to-image diffusion model with neural radiance field (NeRF) through an innovative sampling strategy to incorporates multiple keyframes representing different camera viewpoints and timestamps into a single diffusion model. The edits are made in a canonical space and propagated to remaining time steps through a pretrained deformation network.

In this study, we revisit the issue of geometry editing using a new localized retraining technique that allows for the flawless integration of geometry manipulation tasks from modern 3D modeling software into Neural Radiance Fields. This improves NeRF's suitability as a representation for User Generated Content (UGC) and AI Generated Content (AIGC).

\begin{figure*}[tbp]
    \centering
    \def\svgwidth{2.0\columnwidth}
    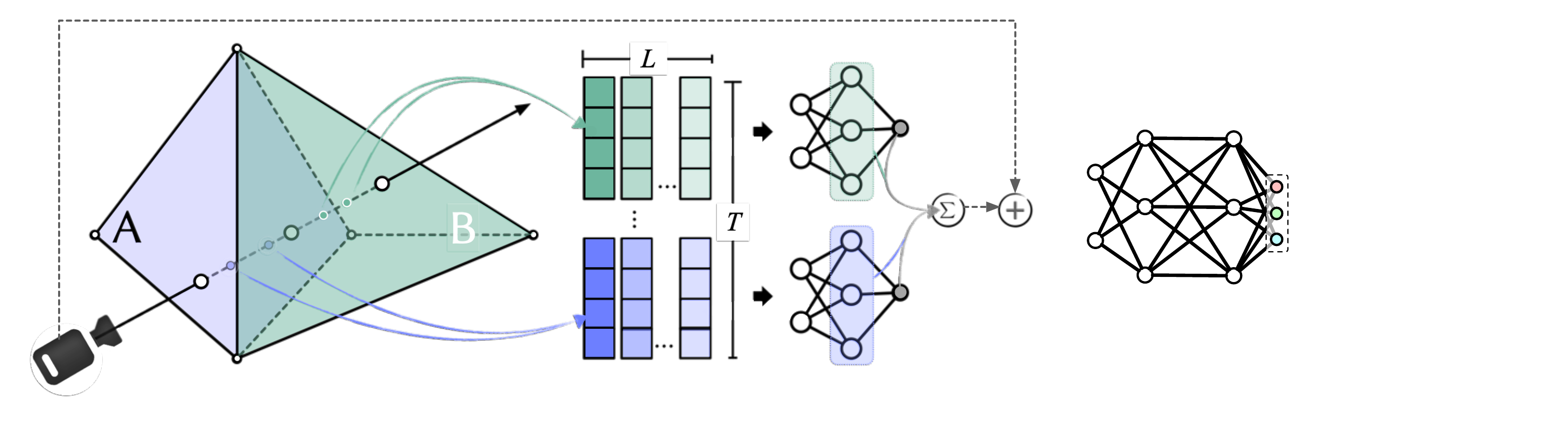

    \caption{Rendering of the \textit{Neural Impostor} starts by determining intersection points between a ray and a tetrahedron mesh, then samples query points between the in and out intersections, based on the size of the tetrahedron (a).
    For each sample point, we use its barycentric coordinates and corresponding tetrahedron to query the hash table (b). A series of small multi-layer perceptrons (MLPs) calculate the spatial density for each sampling point from positional features stored in the hash table (c). To improve rendering efficiency, the early integration process aggregates the features of the sampling points, then alpha blends them into a final appearance feature vector for each ray (d). The final step involves calculating the per-ray RGB color with a substantial appearance MLP and summing over all tetrahedrons (e-f).}
    \label{fig:ni:overview}
\end{figure*}
% \section{Modeling Neural Impostor}
% \label{sec:modeling}
\section{Neural Impostor: A Hybrid Explicit-Implicit Neural Field}
\label{sec:modeling}
\textit{Neural Impostor} uses barycentric coordinates defined by the tetrahedron mesh as the encoding space.
Compared to the Cartesian coordinate-based encoding in the neural radiance field, barycentric coordinates have inherent advantages.
On one hand, barycentric coordinates are symmetrical to the $n+1$ vertices of the $n$-simplex that defines them.
During the vertex-based deformation process, the points within this simplex have a unique barycentric representation.
This allows the \textit{Neural Impostor} to maintain stability during the model's physical deformation process and avoids effects like flickering. On the other hand, barycentric coordinates are intrinsically normalized expressions.
During the encoding process, no additional normalization is needed, providing good support for scenes of any scale.
Simultaneously, barycentric coordinates establish a continuous space subject to linear interpolation. Consequently, \textit{Neural Impostor} transcends resolution limitations, facilitating the high-quality rendering of intricate structures, such as hair.
In the following sections, we will reconstruct the five components of the neural radiance field to elaborate on the modeling method of the \textit{Neural Impostor}.

% \begin{figure*}[t]
%     \centering
%     \includegraphics[width=\textwidth]{figs/pipeline_1.pdf}
%     \caption{Forward rendering process of the \textit{Neural Impostor}. When a ray passes through a tetrahedron mesh, we first calculate the intersection points between the ray and tetrahedron vertices, and sample the implicit field according to the tetrahedron size (a). Then, we obtain the barycentric coordinates of each sampling point (b). Next, we use a multi-layer perceptron (MLP) to obtain the spatial density of each sampling point. To improve rendering efficiency, we first bake the sampling point features into the surface of the tetrahedron, and then decode the color information of each tetrahedron through the MLP. Finally, we complete the rendering through transparency blending (c -> g).}
%     \label{fig:ni:overview}
% \end{figure*}
% \subsection{Barycentric Encoding Space with tetrahedral Meshes}
\subsection{Tetrahedral Neural Representation}
\label{sec:ni:bary_enc}
As shown in \figref{teaser}, to construct an editable implicit radiance field, we reintroduce an explicit representation of its geometric form while modeling the radiance field representing the appearance.
That is, a coarse tetrahedron mesh that acts as the boundary for shape editing, also known as the "cage" \cite{ju2005mean,joshi2007harmonic}.
% Unlike traditional shape editing methods based on tetrahedron cages, \textit{Neural Impostor} has no restrictions on mesh resolution. That is, for complex scenes, we do not need a correspondingly high-resolution mesh, but instead, we adopt the implicit neural field expression of NeRF,% 
Unlike those traditional cage-based shape editing methods, the tetrahedral cage in our NeRF model does \emph{not} have a fine mesh embedded in\textemdash after all, the bare bones of a NeRF model lie in its implicit representation of the radiance field.
Therefore, within each tetrahedron of the cage, we use MLPs to represent its view-dependent radiance field distribution,
maintain a separate radiance field in each tetrahedron of the cage, and convert the position encoding under the original Cartesian coordinate system into the barycentric coordinate space defined by the cage.
At the same time, because the definition of distance and angle under barycentric coordinates is not clear, we retain the original direction encoding method of NeRF under the Cartesian coordinate system, to model the view-dependent effect by introducing the direction of light in the subsequent decoding process (as seen in \figref{ni:overview}).

Specifically, we use $M=(V,T)$ to represent the tetrahedron mesh of the Neural Impostor, where $V=(v_1,v_2,...,v_N)$ represents the set of $N$ vertices of the tetrahedron mesh, and $T=(t_1,t_2,...,t_K)$ represents the $K$ tetrahedron that make up the tetrahedron mesh. Each tetrahedron $t_k\in T$ is represented by its four vertices $V_t={v_0^k,v_1^k,v_2^k,v_3^k}$. For each position coordinate $p$ in Cartesian space, we map it to the corresponding tetrahedron $t$ and the corresponding barycentric coordinates $\Lambda_t={\lambda_0,\lambda_1,\lambda_2,\lambda_3}$ using the barycentric coordinate query function $\mathsf{Q}(\cdot)$ to construct the encoding space, that is:
\begin{equation}
    \begin{aligned}
         & \mathcal{G}_M = \bigcup_{t\in T}\Lambda_t                                                                      \\
         & \Lambda_t= \Bigl\{
        \{\lambda_0^t,\lambda_1^t,\lambda_2^t,\lambda_3^t\}\textit{ }\mid\textit{ }0\leq\lambda_i^t\leq 1\textit{ and }\sum_{i}\lambda_i^t=1
        \Bigl\}                                                                                                           \\
         & \mathsf{Q}: \mathbb{R}^3 \to \mathbb{R}^5, \quad \mathsf{Q}(p)=(t,\Lambda_t),\quad\textit{and } p=\Lambda_tV_t
        \label{eq:ni:tetquery}
    \end{aligned}
\end{equation}
during rendering, $p$ can be transformed to be represented in a ray coordinate system with the ray origin $\dot{o}$ as the coordinate origin and the ray direction $\vec{d}$ as the $z$-axis. Considering the good linear interpolation characteristics of barycentric coordinates, for any point $p$ inside the tetrahedron, we can obtain it by linearly interpolating the barycentric coordinates of the incident and exit points of the ray relative to the tetrahedron, namely:
\begin{equation}
    \begin{aligned}
         & \Lambda_t^{p}=\alpha \Lambda_t^{p_0} + (1-\alpha)\Lambda_t^{p_1},                               \\
         & \textit{where }p_0=\dot{o}+t_0\vec{d},\quad p_1=\dot{o}+t_1\vec{d},\quad\textit{and } t_0 < t_1
        \label{eq:ni:tetsample}
    \end{aligned}
\end{equation}
where $\alpha\in[0,1]$ determines the relative distance of $p$ to the entrance/exit points. Finally, we pack the barycentric coordinates of each sampling point $\Lambda_t^p$, the tetrahedron index $t$, and the light ray direction in the Cartesian coordinate system $\vec{d}$ together to represent a sample in the encoded space $\mathcal{G}$.

Apart from enabling shape editing (see \secref{ni:operations:morphing}), the tetrahedral cage can be also used as a proxy for collision processing. It allows the use of full-fledged collision processing algorithms to detect and resolve collisions between a NeRF model and other objects\textemdash an essential component for bringing NeRF models in a physical simulation.
Last but not least, the tetrahedral cage offers a natural data structure that helps to speed up the image rendering process.
Typically, the NeRF rendering algorithm samples a set of points along each camera ray and accumulates a color value.
With the cage, this process can be implemented more efficiently.

\subsection{Robust Space Sampler}\label{sec:ni:sampling}
NeRF\cite{Mildenhall2020NeRFRS} treats the target scene as a volume space with a certain density, and the physical quantities of discrete sampling points are approximated by integration using body rendering. 
Within a certain limit of the number of samples, in order to accurately model the scene details, we want the sampling points to be concentrated as much as possible in places that contribute to the pixel color to avoid computational waste, i.e., the sampling density should approximate the real density distribution of the scene as much as possible. 
For this reason, in NeRF\cite{Mildenhall2020NeRFRS}, researchers designed a sampling method based on probability distribution, i.e., first obtain the approximate density distribution in space by coarse-grained uniform sampling, then perform fine-grained sampling according to the probability density of coarse-grained sampling points, and fuse the coarse and fine-grained sampling points for the integration process of body rendering. 
This process involves querying the spatial density decoder for both coarse- and fine-grained sampling, and therefore greatly reduces the rendering effectiveness.

Instant-NGP\cite{Mller2022InstantNG} accelerates the rendering process with an explicit spatial occupancy grid that roughly marks the empty and non-empty states of space using binary bits. 
The sampling process then skips over sparse space (i.e., empty space), thereby concentrating the sample points near the surface.
% Sample points are allocated exponentially along rays, rather than uniformly, based on the scale of the scene and by cascading occupancy grids. 
% Using Morton code encoding, queries to the binary occupancy grid during the rendering process can be neglected. 
This method of sampling based on the spatial occupancy grid can greatly speed up the rendering process in NeRF. 
However, this approach also requires the spatial occupancy grid to be updated at a certain frequency during the training process, namely re-querying the MLP to obtain the spatial density of the grid vertices. 
% This process is similar to the forward rendering process and thus has a significant impact on rendering efficiency. 
% When modeling dynamic scenes, adopting this occupancy-field-based sampling approach requires us to constantly update the spatial occupancy grid, greatly affecting rendering efficiency. 
% Furthermore, for scenes with large deformations like shattering, although the sampling method based on barycentric coordinates can ensure the stability of sample point features, if the tetrahedron grid is too small or overly discrete,
Besides, the misalignment between the voxel grid structure defining the occupancy field and the tetrahedron grid
% will be exacerbated. 
% As a result, the sampling based on the spatial occupancy field
% may skip valid sample points or introduce unnecessary ones, causing 
may cause jittering or ghosting artifacts.

Therefore, we designed a sampling method based on barycentric coordinates.
According to the truncation theorem\cite{Mller2022InstantNG}, the sampling step length during light propagation should be proportional to the propagation distance along the ray.
This proportion is determined by the predefined cone angle $\theta$.
Considering that the size of each tetrahedron may change significantly during the deformation process, while the barycentric coordinate distances defined by the incident and outgoing points of the tetrahedron remain relatively stable, we use the interpolation weight $\alpha$ from \eqnref{ni:tetsample} to replace the step length in the Cartesian coordinate system, which means:
\begin{equation}\label{eq:ni:sampling}
    \Delta \alpha_{i+1} = \Delta\alpha_i + \theta\cdot A_i
\end{equation}
where $A_i=\sum_i \Delta\alpha_i$ represents the cumulative sampling distance starting from the incident point of the first valid tetrahedron. Consequently, the sampling process can be described as:
\begin{equation}
    \begin{aligned}
         & \mathsf{S}:\mathbb{R}^6\rightarrow \mathbb{R}^{N\times 8} \\
         & \mathsf{S}(\dot{o}, \vec{d})=\Biggl\{
        \{t,\Lambda_{t},\vec{d}\}_{\times N} \ | \ t\in T, \Lambda_t\in\mathbb{R}^4,0\leq \Lambda_t\leq 1 \textit{ and } \|\Lambda_t\|_1=1\Biggr\}
        \label{eq:ni:sampler}
    \end{aligned}
\end{equation}
where $\dot{o},\vec{d}$ represent the origin and direction of the ray, $N$ stands for the number of sample points for each ray, and $(t,\Lambda_t)$ respectively represent the index of the tetrahedron to which the sample point belongs and its 4-dimentional barycentric coordinates. After sampling, each sample point is represented as a set composed of tetrahedron index, barycentric coordinates, and ray direction.

\subsection{Neural Encoding in a Tetrahedron}
\label{sec:ni:hashing}
% To efficiently encode the neural radiance field in the barycentric coordinate system, we borrowed the multi-resolution hashing encoding method from Instant-NGP\cite{Mller2022InstantNG} to encode the neural radiance field for each tetrahedron. The multi-resolution hashing encoding scheme stores features on multi-grid vertices, where multi-resolution features are packaged into fixed-size hash tables based on multi-grid vertex indices. Each hash table exists independently and without interference at each resolution.

% For a given sample point, its feature is obtained through linear interpolation on the vertices at each grid level. Then, we concatenate the interpolated features from all grid levels and the direction parameters to serve as the final features of the sample point for the decoding phase. Compared to dense grid encoding, multiresolution hash encoding has unique advantages in terms of memory footprint and computational efficiency. On the one hand, the linear storage structure of the hash table is more efficient when performing lookups. On the other hand, as physical quantities such as density and color are relatively smooth in three-dimensional space, we can use a hash table that is much smaller than the size of the dense grid, along with a subsequent MLP decoder to handle hash conflicts, to achieve encoding quality comparable to that of a dense grid. 

In every single tetrahedron, we need to store a view-dependent neural radiance field.
Furthermore, the radiance field must transform smoothly when the tetrahedron is deformed to avoid flickering artifacts.
With this in mind, we represent the positional quantity of the neural radiance field in barycentric coordinates.
As the light ray direction of the neural radiance field is a deformation-independent quantity, we encode the light ray direction in the form of spherical harmonic functions in original Cartesian coordinate system.
The barycentric coordinate system is a local linear space relative to each tetrahedron.
When we change the tetrahedron's vertex positions, points with fixed barycentric coordinates will move correspondingly.
Therefore, under the encoding method of the  \textit{Neural Impostors} based on the tetrahedron barycentric coordinates, when the tetrahedron acting as a proxy deforms, the neural radiance field encoded by each tetrahedron will also undergo corresponding shape changes.
Represent the tetrahedron lookup function as $\mathsf{Q}$, local coordinate encoder inside tetrahedron $t$ as $\mathsf{E}_p^t$ and the global direction encoder as $\mathsf{E}_d$, then the encoding process of the \textit{Neural Impostors} at sampling point $p$ can be represented as:
\begin{equation}
    \begin{aligned}
         & \mathsf{Q}:\mathbb{R}^3\rightarrow\mathbb{I}, \quad t = \mathsf{Q}(p)                                                 \\
         & \mathsf{E}_p^t:\mathbb{R}^4\rightarrow\mathbb{R}^{F\times L}, \quad f_p=\mathsf{E}_p^t (\Lambda_t^p)=h_t(\Lambda_t^p) \\
         & \mathsf{E}_d:\mathbb{R}^3\rightarrow\mathbb{R}^{D^2}, \quad f_d=\mathsf{E}_d(\vec{d})=\textit{SH}_{D}(\vec{d})
        \label{eq:ni:encoder}
    \end{aligned}
\end{equation}
where $\Lambda_t^p$ is the barycentric coordinate of the sampling point $p$ inside the tetrahedron $t$, $L$ is the number of levels in the multiresolution hash, $F$ is the number of features in each level, $D$ is the order of the spherical harmonic function encoding, $h_t$ represents the feature querying processing from tetrahedron $t$'s local hash table, and $\textit{SH}_D$ is the $D$-order spherical harmonic function. After encoding, we can obtain the position feature $f_p$ and direction feature $f_d$ corresponding to each sampling point $p$.

However, the use of barycentric coordinates complexes the encoding process (i.e., $\mathsf{E}_p^t$ in \eqnref{ni:encoder}). We wish to leverage the multi-resolution hash encoding scheme proposed in ~\cite{Mller2022InstantNG}, as it offers
high rendering quality and low memory footprint. This encoding
scheme conceptually stores trainable features on a multi-resolution Cartesian grid, and directly transferring this encoding scheme in barycentric coordinates requires the creation of a multi-resolution tetrahedron mesh within a tetrahedron and the storage of feature vectors on the tetrahedron mesh's vertices.  Yet, this approach is troublesome. When a tetrahedron is subdivided for finer level mesh creation, it results in four smaller tetrahedron and an octahedron
in the center. It is unclear how to split the octahedron into a set of tetrahedron in a consistent way.  More importantly, when it comes to image rendering, a recurring step is to find, on each multi-resolution level, the tetrahedron in which a sampled point on a camera ray is located. This, however, is computationally much more expensive than its counterpart in \cite{Mller2022InstantNG} (i.e. finding in an axis-aligned Cartesian grid the voxel in which a sampled point is located).

% \begin{figure}[t]
%     \centering
%     \includegraphics[width=\linewidth]{figs/tet_subdivision.pdf}
%     \vspace{-2mm}
%     \caption{Schematic diagram of tetrahedron subdivision. Based on the midpoints of the original tetrahedron edges, we can divide it into four smaller tetrahedron and one central octahedron.}
%     \label{fig:tet_subdivision}
%     \vspace{-2mm}
% \end{figure}

Instead, we propose to maintain trainable features not on any tetrahedral vertices but on a four-dimensional (4D) grid associated with the tetrahedron.
The 4D grid is multiresolution, constructed in the following way.
Consider a point $p$ inside the tetrahedron $t$ with barycentric coordinates $\Lambda_t(p)={\lambda_0^t,\lambda_1^t,\lambda_2^t,\lambda_3^t}$. Although its barycentric coordinates only have three degrees of freedom (i.e., constrained by $\sum_{i=0}^3\lambda_i^t=1$), algebraically it can be regarded as a 4D point located inside a 4D voxel.
This 4D voxel has 16 vertices (corners), each with a coordinate $(u_1,u_2,u_3,u_4)\in{0,1}^4$.
This voxel forms the highest-level grid with a resolution of $1\times1\times1\times1$.
We create the multiresolution grid by progressively subdividing this highest-level grid, and bind the trainable features characterizing the radiance field of the tetrahedron to the nodes of the multiresolution grid.

\begin{figure}[t]
    \centering
    \def\svgwidth{1\columnwidth}
    %% Creator: Inkscape 1.2.2 (b0a8486541, 2022-12-01), www.inkscape.org
%% Accompanies image file 'tet_encoding.pdf' (pdf, eps, ps)
\begingroup%
  \makeatletter%
  \providecommand\color[2][]{%
    \errmessage{(Inkscape) Color is used for the text in Inkscape, but the package 'color.sty' is not loaded}%
    \renewcommand\color[2][]{}%
  }%
  \providecommand\transparent[1]{%
    \errmessage{(Inkscape) Transparency is used (non-zero) for the text in Inkscape, but the package 'transparent.sty' is not loaded}%
    \renewcommand\transparent[1]{}%
  }%
  \providecommand\rotatebox[2]{#2}%
  \newcommand*\fsize{\dimexpr\f@size pt\relax}%
  \newcommand*\lineheight[1]{\fontsize{\fsize}{#1\fsize}\selectfont}%
  \ifx\svgwidth\undefined%
    \setlength{\unitlength}{2400bp}%
    \ifx\svgscale\undefined%
      \relax%
    \else%
      \setlength{\unitlength}{\unitlength * \real{\svgscale}}%
    \fi%
  \else%
    \setlength{\unitlength}{\svgwidth}%
  \fi%
  \global\let\svgwidth\undefined%
  \global\let\svgscale\undefined%
  \makeatother%
  \begin{picture}(1,1)%
    \lineheight{1}%
    \setlength\tabcolsep{0pt}%
    \put(0.23524009,0.49706622){\color[rgb]{0,0,0}\makebox(0,0)[lt]{\lineheight{1.25}\smash{\begin{tabular}[t]{l}(a)\end{tabular}}}}%
    \put(0.73372533,0.49706622){\color[rgb]{0,0,0}\makebox(0,0)[lt]{\lineheight{1.25}\smash{\begin{tabular}[t]{l}(b)\end{tabular}}}}%
    \put(0.23640993,0.01153268){\color[rgb]{0,0,0}\makebox(0,0)[lt]{\lineheight{1.25}\smash{\begin{tabular}[t]{l}(c)\end{tabular}}}}%
    \put(0.73372533,0.01153268){\color[rgb]{0,0,0}\makebox(0,0)[lt]{\lineheight{1.25}\smash{\begin{tabular}[t]{l}(d)\end{tabular}}}}%
    \put(0,0){\includegraphics[width=\unitlength,page=1]{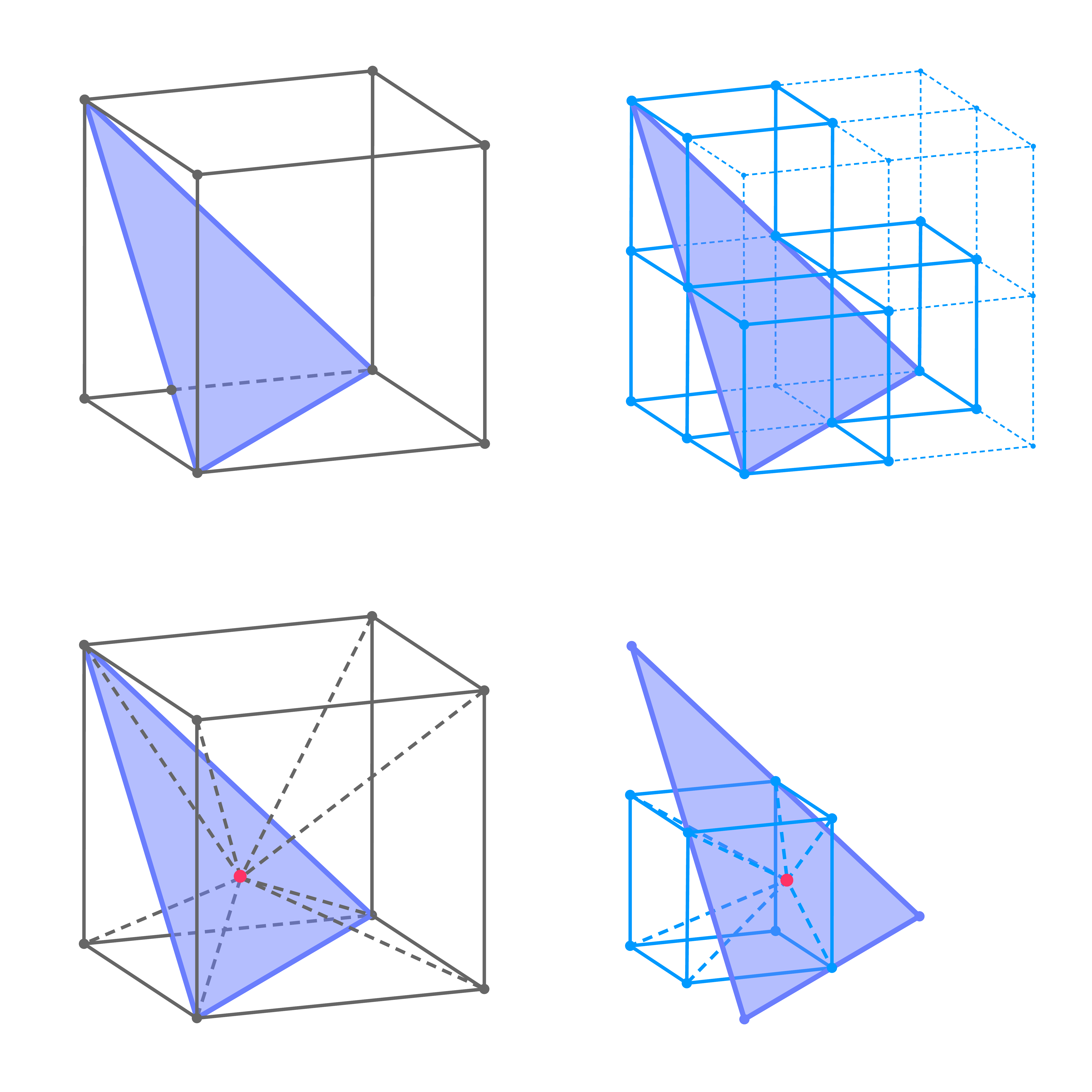}}%
  \end{picture}%
\endgroup%

    \vspace{-2mm}
    \caption{Multi-grid hash encoding for a triangle, its barycentric coordinates define a three-dimensional encoding space, but the valid encoding space is merely a two-dimensional embedding (purple triangle) within this three-dimensional space. (a) and (c) gives the valid voxels and voxel corners used for interpolation under the first grid level, while (b) and (d) demonstrate the corresponding scenario after subdivision.}
    \label{fig:4d_vis}
\end{figure}

The relation of a tetrahedron to its 4D grid can be interpreted geometrically.
On the highest-level grid\textemdash which is a single 4D voxel\textemdash there are 16 voxel
vertices. Four of them have coordinates satisfying $\sum_{i=1}^4u_i=1$,
which are valid barycentric coordinates corresponding to the tetrahedron's four
vertices.  In light of this, the tetrahedron can be viewed as a 3D region
embedded in a 4D hyperspace.  A sampled point $p$ on a camera ray is always located
in the tetrahedron, but in order to compute the feature vector
at $p$, we interpolate the features stored in the corners of a 4D voxel (see \figref{4d_vis}).
Further, the highest-level grid is subdivided into 16 4D voxels on the second
level of the grid. But the 3D tetrahedron is embedded in only 5 of the 16 voxels.
The rest of the voxels remains unused when we compute feature vectors of
sampled ray points.
See \figref{4d_vis} for visualization of this interpretation in 3D space.

We note that the unused voxels will not waste memory storage, because
following Instant-NGP\cite{Mller2022InstantNG}, the feature vectors on grid nodes
are stored in a hash table, and the hash table size is set on purpose much
smaller than the grid size to ``compress'' feature vector storage.
Meanwhile, considering that we need to maintain separate hash tables for each tetrahedron in the tetrahedron mesh of the \textit{Neural Impostor}, when the number of tetrahedrons increases, the amount of information each tetrahedron actually needs to encode also decreases. Therefore, in practice, we reduce the hash table size of each tetrahedron according to the number of tetrahedrons in the mesh, in order to maintain the hash table size of the entire tetrahedron mesh comparable to that in the original Instant-NGP with a given level. For instance, consider a tetrahedron mesh with $T$ tetrahedrons and base-2 logarithmic hash table size $H$ , the corresponding local base-2 logarithmic hash table size in each tetrahedron would be $H'=H / \lfloor(\log_{2}T+1)\rfloor$. The rest of the encoding process is similar to Instant-NGP.
At each multiresolution grid level $l$, the feature vector $\boldsymbol{f}_l(\boldsymbol{x})$ at the sampling point $\boldsymbol{x}$ is obtained through four-dimensional linear interpolation, using the feature vectors at the $16$ corner points of the voxel containing $\boldsymbol{x}$. We concatenate feature vectors $\boldsymbol{f}_l(\boldsymbol{x})$ from all levels $l=1...L$ to form the feature vector $\boldsymbol{f}_p$ used for decoding. The decoding process of \textit{Neural Impostor} mirrors that of NeRF\cite{Mildenhall2020NeRFRS}. Given a density decoder, denoted $\mathsf{D}_{\sigma}$, and a radiance decoder $\mathsf{D}_{r}$, we have:
\begin{equation}
    \begin{aligned}
         & \mathsf{D}_{\sigma}:\mathbb{R}^{F\times L}\rightarrow \mathbb{R},\quad \sigma(p)=\mathsf{D}_\sigma(f_p)              \\
         & \mathsf{D}_r: \mathbb{R}^{D^2+F\times L}\rightarrow \mathbb{R}^3,\quad \boldsymbol{c}(p)=\mathsf{D}_r(f_p\oplus f_d)
        \label{eq:ni:decoder}
    \end{aligned}
\end{equation}
herein, $p=\dot{o}+t\vec{d}$ denotes a sampled point along the ray, $f_p$ and $f_d$ represent the position feature and the direction feature of $p$ and $\oplus$ indicates feature concatenation.

\subsection{Efficient Image Rendering}
\label{sec:ni:rendering}
During rendering, we need to calculate the color value $C(\boldsymbol{r})$ along the sampled camera ray $\boldsymbol{r}(t)=\dot{o}+t\cdot\vec{d}$, where $\dot{o}$ is the origin of the ray and $\vec{d}$ is the ray direction. Similar to the original NeRF model~\cite{Mildenhall2020NeRFRS}, following the principles of Direct Volume Rendering (DVR), the color of the ray $C(\boldsymbol{r})$ can be estimated by integrating over a bounded distance interval $[t_n,t_f]$:
\begin{equation}\label{eq:ni:volume_render}
    C(\boldsymbol{r})=\int_{t_n}^{t_f} T(t)\sigma(\boldsymbol{r}(t))\boldsymbol{c}(\boldsymbol{r}(t),\vec{d})\d{t},
\end{equation}
here, $T(t) = \exp\left(-\int_{t_n}^t \sigma(\boldsymbol{r}(s))\d{s}\right)$ represents the accumulated transmittance from $t_n$ to $t$ along the ray direction. $\sigma(\boldsymbol{r})$ is the volume density at location $\boldsymbol{r}(t)$; $\boldsymbol{c}(\boldsymbol{r}(t),\vec{d})$ is the radiance at $\boldsymbol{r}(t)$ in the direction $\vec{d}$.

The fundamental idea of NeRF\cite{Mildenhall2020NeRFRS} is to represent $\sigma(\boldsymbol{r}(t))$ and $\boldsymbol{c}(\boldsymbol{r}(t),\vec{d})$ using neural networks. In our tetrahedron neural representation, the camera ray $\boldsymbol{r}$ intersects with a series of tetrahedron. During the deformation process, the volume of each tetrahedron might undergo significant changes. Under these circumstances, if we keep the Cartesian coordinate-based integral method, the final color and transparency of the ray would directly correlate with the integral distance. Consequently, any changes in the integral distance due to the deformation of the tetrahedron would also affect the output color.

To dissociate the color of the ray from the integral distance, we leverage the sampling method discussed in \secref{ni:sampling}. We rewrite \eqnref{ni:volume_render} as a sum of the integrals for each tetrahedron, while converting the integral distance into a distance representation in barycentric coordinates. This ensures that the color of the ray remains unaffected, even when the volume of the tetrahedron experiences significant deformation. As illustrated in \figref{ni:overview}, we denote the $K$ tetrahedron that the ray passes through during propagation as $t_1,t_1,\cdots,t_K$. Each tetrahedron intersects with the ray at entry point $p_0$ and exit point $p_1$, with corresponding barycentric coordinates $\Lambda_t^0$ and $\Lambda_t^1$, respectively. We use the Manhattan distance between $\Lambda_t^0$ and $\Lambda_t^1$ as the integral distance, and recast \eqnref{ni:volume_render} as a sum of integrals over each tetrahedron:
\begin{equation}
    C(\boldsymbol{r}) = \sum_{i=1}^K\int_{\Lambda_{t_i}^0}^{\Lambda_{t_i}^1} T(\Lambda_{t_i}^p)\sigma(\Lambda_{t_i}^p)\boldsymbol{c}(p,\vec{d})\ d{\Lambda_{t_i}^p}
\end{equation}
Here, $\Lambda_{t_i}^p$ represents the barycentric coordinates of the sampling point $p$ within the tetrahedron $t_i$. Due to the linearity of the barycentric coordinates, $\Lambda_{t_i}^p$ can be determined by the barycentric coordinates of the entry and exit points, $\Lambda_t^0$ and $\Lambda_t^1$, respectively, as well as the interpolation weight $\alpha$ as defined in \eqnref{ni:tetsample}.

The radiance in a scene typically carries a higher frequency of information than the spatial density. Therefore, the radiance decoder in the process of constructing a neural radiance field usually exhibits more complexity than the spatial density decoder. Moreover, the forward process of a MLP is more costly in terms of time efficiency compared to encoding and rendering. To further optimize the rendering efficiency, we first decode the position features of the sample points into density values using a few density decoders $\mathsf{D}_{\sigma}$. In contrast to previous rendering schemes that decode before integration for radiance, we propose a rendering scheme that integrates before decoding. Specifically, we first integrate the position features of the sampling points and ray transmittance onto its belonging tetrahedron surfaces, then alpha composite them together into a final appearance feature for each ray. Subsequently, the composited appearance feature are combined with the ray directional feature for radiance decoding, as such we only need to computer the appearance decoder once for each ray:
\begin{equation}
    C(\boldsymbol{r}) = \mathsf{D}_r\Bigl[ \sum_{i=1}^K 
    \underbrace{\int_{\Lambda_{t_i}^0}^{\Lambda_{t_i}^1} T(\Lambda_{t_i}^p)\sigma(\Lambda_{t_i}^p)
    \mathsf{E}_p^{t_i}(\Lambda_{t_i}^p)\ d{\Lambda_{t_i}^p}}_{\textit{integrated positional feature}} \ \oplus \ \mathsf{E}_d(\vec{d})
    \Bigr]
    \label{eq:ni:tetnerf_render}
\end{equation}
here, $\mathsf{D}_r$ is the radiance decoder, and $\mathsf{E}_p, \mathsf{E}_d$ are the encoders for the barycentric coordinates and ray direction, respectively. The density values of the sample points are obtained by decoding the barycentric coordinates according to \eqnref{ni:decoder}. This "integration-before-decoding" rendering approach effectively pre-bakes the scalar field of the encoding space onto the surface of each tetrahedron. As each ray can define a unique surface point for each tetrahedron, this "integration-before-decoding" approach allows us to rasterize the tetrahedron in the space layer by layer and project them onto the imaging plane to further accelerate the rendering process during inference.

While the "integration-before-decoding" rendering approach can accelerate the rendering process during inference, in the barycentric coordinate-based encoding method, we also need to determine the tetrahedron to which each sample point belongs during the training process. For a 1080p image, the forward process of training requires conducting a barycentric coordinate test for approximately $1080 \times 1080 \approx 10^8$ points. Traditional tests based on matrix multiplication would require us to compute the determinant of a $N_{rays}\times N_{points}\times 4\times 3$ matrix, which is unfeasible for complex tetrahedron meshes in one forward pass. Building upon the work of Wald et al. (2019) \cite{Wald2019RTXBR}, we consider the triangles that make up the tetrahedron as the testing units during the rendering process. For non-self-intersecting tetrahedron meshes, the concept of a point being "in front of a triangle" or "behind a triangle" can be uniquely determined by the winding order of the triangle vertices. In their method, determining to which tetrahedron a point belongs is equivalent to casting a ray from that point in a random direction, evaluating the winding order at the first intersection point, and checking on which side of the triangle the sample lies. Since each triangle belongs to at most two tetrahedron, this winding order-based testing approach can yield a unique tetrahedron index for any point in space. When hardware acceleration is available, the triangle-based query operation is also very efficient in terms of memory and computation.

\begin{figure}
    \centering
    \def\svgwidth{1\columnwidth}
    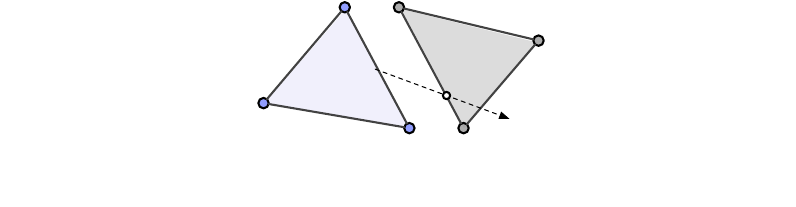

    \caption{Three tetrahedron connection scenarios observed during sampling. For adjacent tetrahedrons sharing a triangle face, the entry point $p_0$ originates the ray, and steps determine the exit point $p_1$, which then updates to the new entry point. For separate tetrahedron, we link them at runtime, moving a point at $p_2$ to $p_3$ before sampling. With intersecting tetrahedron, an aggressive sampling approach adds a point $p_4$ that belongs to both tetrahedron $A$ and $B$ to the sampling set.}
    \label{fig:ni:tet_cases}
\end{figure}

We extend the method of \cite{Wald2019RTXBR} to the framework of ray tracing to support the sampling process in \secref{ni:sampling}. Specifically, in the context of ray tracing, we can perform stepwise sampling along the direction of the ray. Unlike the separate testing for each point in the method of Wald et al. (2019) \cite{Wald2019RTXBR}, we only need to determine the entry and exit points of the tetrahedron that each ray passes through. Depending on the arrangement of tetrahedron in space, the progression of the ray may encounter three situations shown in \figref{ni:tet_cases}, namely adjacent/non-adjacent tetrahedron and intersecting tetrahedron. We adopt different stepping strategies for these three situations. As shown in \figref{ni:tet_cases}, for adjacent tetrahedron that share one triangle faces, we take the entry point as the origin of the ray and step to determine the exit point. After sampling, the entry point is updated to the exit point. For separated tetrahedron, we connect them at runtime. That is, if the sample point is at position $p_2$ in \figref{ni:tet_cases}, we first update it to $p_3$ before sampling. For intersecting tetrahedron, we use an aggressive sampling approach, i.e., for a sample point $p_4$ that belongs to both tetrahedron $A$ and $B$, we add its position in both $A$ and $B$ to the sampling set.

\paragraph*{Rendering Acceleration}
Besides the raymaching-based rendering algorithm stated in \secref{ni:rendering}, we provide a tailored algorithm for ultra-fast rendering in large-scale and multi-object scenes.
We achieve the ultra-fast rendering by baking the appearance of a tetrahedron onto it's faces as view-dependent neural textures.
For each image in the input image set, we cast a ray from the camera through each pixel.
As shown in \figref{ni:overview} (a), we first compute the intersection points between the ray and the tetrahedral mesh envelope, using the NVIDIA Optix library for this computation.
Then, along each ray, we sample the implicit field and integrate the feature values of the sampled points onto the tetrahedral surface using an early integration approach.
Finally, we obtain the pixel color through opacity blending.
During inference, we can bake the volumetric features of each tetrahedron onto the tetrahedral surface and render the scene using rasterization.
\tabref{ni:query_speed} provides the efficiency comparison between our implementation and the barycentric coordinate query-based rendering approach used in NeRF-Editing \cite{nerf-editing2022} with \textit{KNN} neighborhood query. 
% (Due to memory limitations, NeRF-Editing\cite{nerf-editing2022} was tested at a resolution of $240\times 180$).
As seen, even without baking, we can achieve real-time volume rendering with the hardware acceleration of Optix (method "Optix Accelerated" in \tabref{ni:query_speed}).
After baking, rendering efficiency can be significantly improved by using rasterization-based rendering in a multi-layer fashion (method "Baking+Rasterization" in \tabref{ni:query_speed}).

\section{Editing operations of Neural Impostor}
\label{sec:operations}
A comprehensive explicit geometric editing framework typically includes continuous editing based on spatial transformations, and topology editing which changes the topological structure of the shape. More specifically, we divide editing operations into three categories: continuous shape morphing, explicit remeshing, and boolean operations. Continuous shape morphing changes only the positions of the vertices of the explicit tetrahedron mesh without altering its topological structure. Explicit remeshing regenerates the explicit mesh to enable detailed editing. Boolean operations are used to logically combine two or more shapes to create a new object.

In addition to supporting basic geometric editing operations, the hybrid modeling approach based on \textit{Neural Impostor} allows complex combinations of operations on implicit fields.
By utilizing the aforementioned operations,
our system achieves a wide range of editing tasks, including physical simulation, mesh reconstruction, and scene composition.
In the following, we will explain how we leverage explicit tetrahedron meshes in \textit{Neural Impostor} to realize the aforementioned editing operations for neural implicit field.

\begin{figure*}[t]
    \centering
    \def\svgwidth{2.0\columnwidth}
    %% Creator: Inkscape 1.2.2 (b0a8486541, 2022-12-01), www.inkscape.org
%% Accompanies image file 'shape_morphing.pdf' (pdf, eps, ps)
\begingroup%
  \makeatletter%
  \providecommand\color[2][]{%
    \errmessage{(Inkscape) Color is used for the text in Inkscape, but the package 'color.sty' is not loaded}%
    \renewcommand\color[2][]{}%
  }%
  \providecommand\transparent[1]{%
    \errmessage{(Inkscape) Transparency is used (non-zero) for the text in Inkscape, but the package 'transparent.sty' is not loaded}%
    \renewcommand\transparent[1]{}%
  }%
  \providecommand\rotatebox[2]{#2}%
  \newcommand*\fsize{\dimexpr\f@size pt\relax}%
  \newcommand*\lineheight[1]{\fontsize{\fsize}{#1\fsize}\selectfont}%
  \ifx\svgwidth\undefined%
    \setlength{\unitlength}{2639.00006104bp}%
    \ifx\svgscale\undefined%
      \relax%
    \else%
      \setlength{\unitlength}{\unitlength * \real{\svgscale}}%
    \fi%
  \else%
    \setlength{\unitlength}{\svgwidth}%
  \fi%
  \global\let\svgwidth\undefined%
  \global\let\svgscale\undefined%
  \makeatother%
  \begin{picture}(1,0.31186052)%
    \lineheight{1}%
    \setlength\tabcolsep{0pt}%
    \put(0,0){\includegraphics[width=\unitlength,page=1]{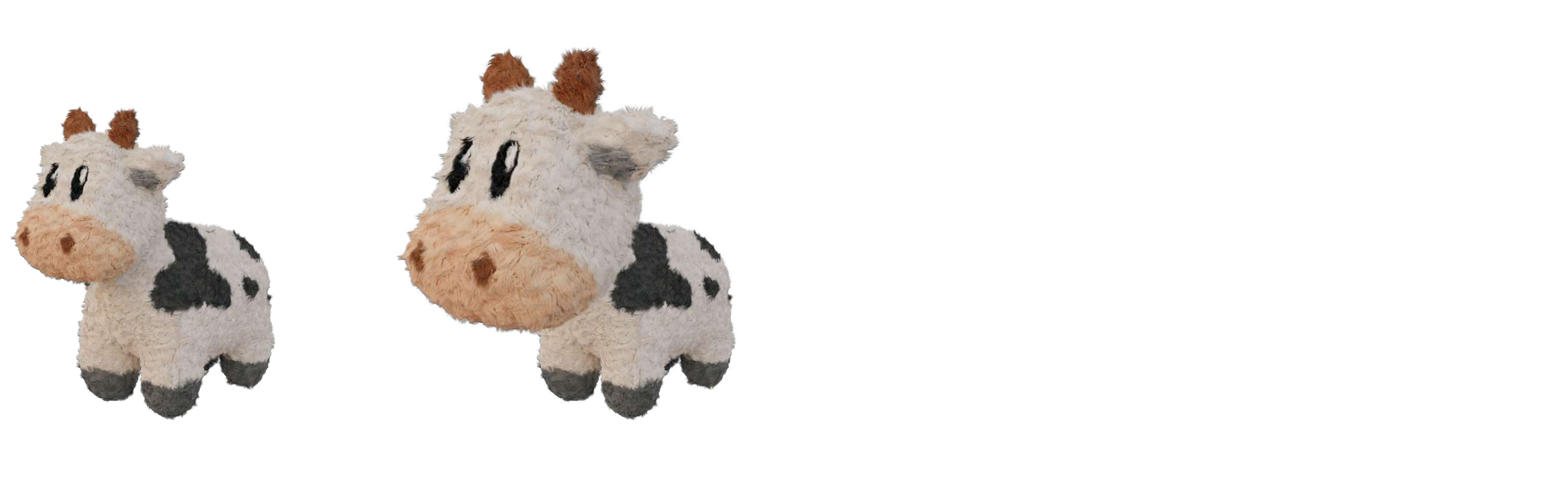}}%
    \put(0.04640576,0.01136793){\color[rgb]{0,0,0}\makebox(0,0)[lt]{\lineheight{1.25}\smash{\begin{tabular}[t]{l}(a) Original\end{tabular}}}}%
    \put(0,0){\includegraphics[width=\unitlength,page=2]{shape_morphing.pdf}}%
    \put(0.56456337,0.01136793){\color[rgb]{0,0,0}\makebox(0,0)[lt]{\lineheight{1.25}\smash{\begin{tabular}[t]{l}(c) Rotate Head\end{tabular}}}}%
    \put(0,0){\includegraphics[width=\unitlength,page=3]{shape_morphing.pdf}}%
    \put(0.29781715,0.01136793){\color[rgb]{0,0,0}\makebox(0,0)[lt]{\lineheight{1.25}\smash{\begin{tabular}[t]{l}(b) Zoom Head\end{tabular}}}}%
    \put(0,0){\includegraphics[width=\unitlength,page=4]{shape_morphing.pdf}}%
    \put(0.80711212,0.01136793){\color[rgb]{0,0,0}\makebox(0,0)[lt]{\lineheight{1.25}\smash{\begin{tabular}[t]{l}(d) Move Horns\end{tabular}}}}%
    \put(0,0){\includegraphics[width=\unitlength,page=5]{shape_morphing.pdf}}%
  \end{picture}%
\endgroup%

    \caption{Shape morphing based on vertices with local remeshing. In this process, we scale some vertices of (a) the original tetrahedron mesh to achieve (b) local scaling and (c) the rotating motion of the neural radiance field. Further, (d) we remesh the ear region for finer scale shape morphing.}
    \label{fig:ni:morphing}
\end{figure*}
\subsection{Shape Transformation and Deformation}
\label{sec:ni:operations:morphing}
Shape transformation and deformation are the most fundamental geometric editing operations.
Fundamentally, shape transformation and deformation reveal geometric invariance under continuous mappings, which can naturally be satisfied by the hybrid modeling of \textit{Neural Impostor}.
Thanks to the centroid coordinate encoding method described in \secref{ni:bary_enc}, in \textit{Neural Impostor} any spatial transformation based on tetrahedron vertices can be naturally transformed into the implicit field.
For example, in \figref{ni:morphing}, the input shape morphs from the original shape $M$ on the left to the head rescaled scaled shape $M'$ on the right.
Recalling equation \ref{eq:ni:tetquery}, the sampled points in the deformed tetrahedron mesh $M'$ are given by:
\begin{equation}
    t, \Lambda_t^p\leftarrow \mathsf{S}(p, M'), \textit{ where }p\in \mathbb{R}^3\text{\textit{ and }} t\in M'
\end{equation}

In this case, both the sampling spaces before and after deformation are defined in the Cartesian space of the scene.
When the tetrahedron vertices move, the implicit neural field inside the deformed tetrahedron remain unchanged under the barycentric coordinate.
Thus, we can adopt the same sampling, encoding, and rendering methods as the original process without any additional transformations.
Moreover, as this process only involves updates to the tetrahedron mesh vertices without calculating the transformations on sampling points, the deformation process in \textit{Neural Impostor} is more efficient than approaches based on bending rays\cite{cage2022,Peng2022CageNeRF,voltemorph2022,nerf-editing2022}.

\subsection{Subdivision and Remeshing}
Compared to continuous shape morphing, topological operations pose more challenges. When the mesh topology changes, the number and connectivity of the original tetrahedron are altered. However, in \textit{Neural Impostor} the radiance field is strictly tied to the tetrahedron topology.
This requires an effective mapping scheme to transfer the radiance field baked into the original tetrahedron mesh to the new tetrahedron structure.
% Under the traditional \textit{Neural Radiance Fields} (NeRF) encoding scheme, when there is a change in the encoding space of the neural radiance field, such as when the grid space or tetrahedron mesh moves, it requires retraining the implicit field using a deformed image set as supervision.

\begin{figure*}[t]
    \centering
    \def\svgwidth{2\columnwidth}
    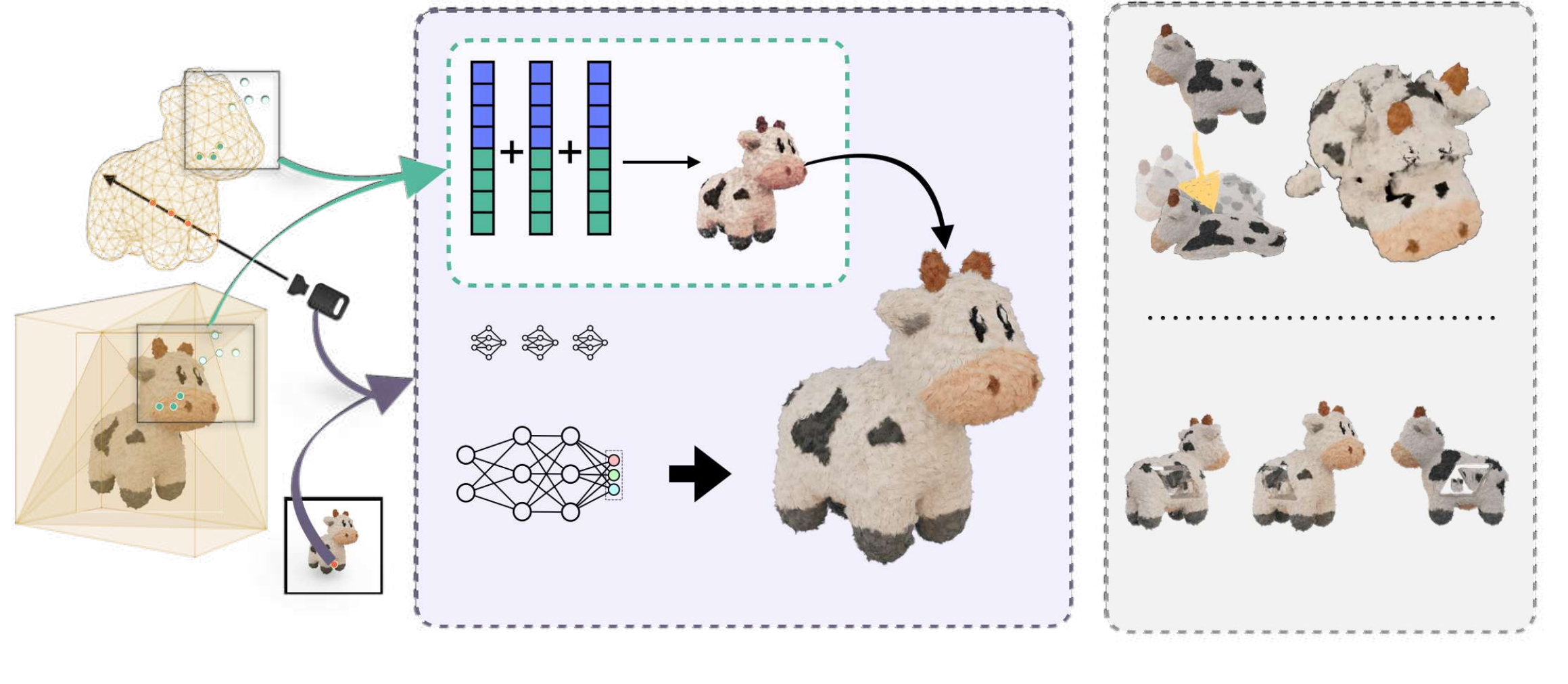

    \caption{Two-Stage Retraining Process. During (c) Stage 1, the hashtable is optimized using randomly sampled points to align with the existing \textbf{ Neural Impostor}. Only points located within the updated proxy (solid green points) are taken as valid samples, while others (hollow points with green borders) are disregarded. However, solely updating the hash table could potentially result in noisy rendering due to hash collision. To mitigate this issue, a 2nd-stage training (d) is introduced to refine the decoder MLPs and resolve hash collision. Here, rays are randomly traced into the updated proxy, with the accumulated per-ray color further guiding the adjustment of density and color decoders for the new \textbf{Neural Impostor}. This efficient local retraining mechanism facilitates (e) fine-scale deformation and (f) local appearance editing.}
    \label{fig:ni:remesh}
\end{figure*}

% However, for edited implicit fields, the ground truth image set is undefined. In the representation of \textit{Neural Impostor} deformation and scene appearance are decoupled in a true sense.
% This allows us to use a new topology to transfer the implicit neural field defined on the original topology before deformation, achieving consistent visual effects during topological transformations.
To achieve this goal, we propose a local retraining strategy to synthesize neural implicit fields from existing neural implicit fields with different explicit tetrahedral mesh.
% When the tetrahedron mesh topology changes, the implicit field of the new topology can be obtained through the local retraining of the implicit field defined on the original geometry.
When remeshing happens, rather than performing a global retraining, we select local tetrahedron where the topology changes for retraining while keeping the other tetrahedron unchanged.
% In other words, instead of performing global retraining, we select local tetrahedron where the topology changes for retraining while keeping the other tetrahedron unchanged.
Besides, to accelerate convergence, we initialize the updated multigrid hash table based on feature-only supervision shown in\figref{ni:remesh} (c).
Then follow a joint optimization on both the feature hash table and the MLP weights.
% Then, if the real image set is available, we can further optimize the results using synthesized colors as supervision (\figref{ni:remesh} (d)).
As shown in \figref{ni:remesh}, during the retraining stage, our input is the pre-trained \textit{Neural Impostor} which includes its tetrahedron mesh and the implicit neural field encoded in each tetrahedron illustrated in \figref{ni:remesh} (b), as well as the explicit tetrahedron mesh after the topological transformation as shown in \figref{ni:remesh} (a).
The goal of local retraining is to transfer the radiance field defined on the original tetrahedrons to the new radiance field on the remeshed tetrahedral mesh.
% In the retraining process, we require the reconstructed mesh (a) to cover a similar effective region as the original mesh, which can be achieved by deforming the vertices of the transformed mesh to some extent.
The retraining process consists of two stages. In the first stage, as shown in \figref{ni:remesh}(c), we randomly sample points inside (solid dots in \figref{ni:remesh} (a)) and outside (hollow dots in \figref{ni:remesh} (a)) of the new mesh to optimize the hash table. In the second stage (\figref{ni:remesh} (d)), we use ray tracing to locate the pixel positions in that region and select effective rays to obtain pixel colors as further supervision according to the rendering process in \eqnref{ni:tetnerf_render}, this improving the accuracy of the rendered results after retraining.
Specifically, let $(M, h, \mathsf{D})$ represent the tetrahedron mesh, hash encoding table, and decoder of the \textit{Neural Impostor} before retraining, and let $(M', h', \mathsf{D}')$ represent the reconstructed tetrahedron mesh and its to-be-trained hash encoding table and decoder. Let $\mathsf{R}$ and $\mathsf{E}_d$ represent the renderer and ray direction encoder used in the second stage training. The training process can be formalized as follows:
\begin{equation}
    \begin{aligned}
        \textit{Stage 1: } & \mathcal{L}_{hash}=\|h(\Lambda_p)-h'(\Lambda_p)\|^1, \textit{ where }p\in M\cup M'. \\
        \textit{Stage 2: } & \mathcal{L}_{render}=\Bigg\|\mathsf{R}\circ\mathsf{D}
        \bigl[ h(\Lambda_p)\oplus \mathsf{E}_d(\vec{d})\bigr]-\mathsf{R}\circ\mathsf{D'}
        \bigl[ h'(\Lambda_p)\oplus \mathsf{E}_d(\vec{d})\bigr]\Bigg\|^2,                                         \\
                           & \textit{where }p\in\dot{o}+t\vec{d}.
    \end{aligned}
\end{equation}
where $\circ$ denotes function composition and $\oplus$ represents feature concatenation.

\subsection{Boolean Operation}
\begin{figure*}[t]
    \centering
    \def\svgwidth{2\columnwidth}
    %% Creator: Inkscape 1.2.2 (732a01da63, 2022-12-09), www.inkscape.org
%% Accompanies image file 'boolean_op.pdf' (pdf, eps, ps)
\begingroup%
  \makeatletter%
  \providecommand\color[2][]{%
    \errmessage{(Inkscape) Color is used for the text in Inkscape, but the package 'color.sty' is not loaded}%
    \renewcommand\color[2][]{}%
  }%
  \providecommand\transparent[1]{%
    \errmessage{(Inkscape) Transparency is used (non-zero) for the text in Inkscape, but the package 'transparent.sty' is not loaded}%
    \renewcommand\transparent[1]{}%
  }%
  \providecommand\rotatebox[2]{#2}%
  \newcommand*\fsize{\dimexpr\f@size pt\relax}%
  \newcommand*\lineheight[1]{\fontsize{\fsize}{#1\fsize}\selectfont}%
  \ifx\svgwidth\undefined%
    \setlength{\unitlength}{1095bp}%
    \ifx\svgscale\undefined%
      \relax%
    \else%
      \setlength{\unitlength}{\unitlength * \real{\svgscale}}%
    \fi%
  \else%
    \setlength{\unitlength}{\svgwidth}%
  \fi%
  \global\let\svgwidth\undefined%
  \global\let\svgscale\undefined%
  \makeatother%
  \begin{picture}(1,0.52694064)%
    \lineheight{1}%
    \setlength\tabcolsep{0pt}%
    \put(0,0){\includegraphics[width=\unitlength,page=1]{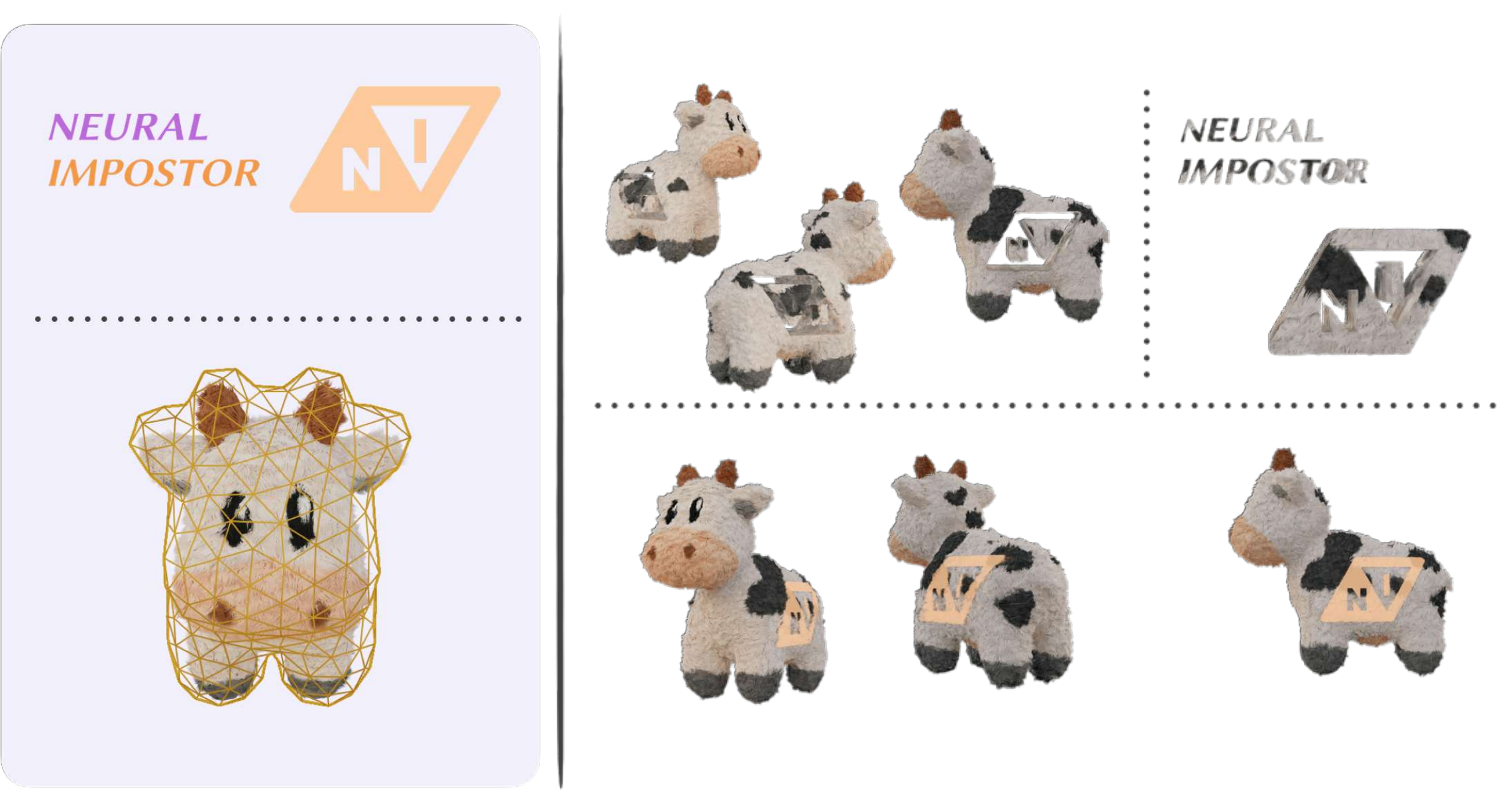}}%
    \put(0.17697098,0.3316648){\color[rgb]{0,0,0}\makebox(0,0)[t]{\lineheight{1.25}\smash{\begin{tabular}[t]{c}(a) Pattern Brushes\end{tabular}}}}%
    \put(0.17656073,0.02263317){\color[rgb]{0,0,0}\makebox(0,0)[t]{\lineheight{1.25}\smash{\begin{tabular}[t]{c}(b) Existing Neural Impostor\end{tabular}}}}%
    \put(0.69488208,0.49049266){\color[rgb]{0,0,0}\makebox(0,0)[t]{\lineheight{1.25}\smash{\begin{tabular}[t]{c}(c) Boolean Operations\end{tabular}}}}%
    \put(0.69524773,0.02263317){\color[rgb]{0,0,0}\makebox(0,0)[t]{\lineheight{1.25}\smash{\begin{tabular}[t]{c}(d) Boolean-based Appearance Editing\end{tabular}}}}%
  \end{picture}%
\endgroup%

    \caption{Boolean operations on \textit{Neural Impostor} with given patterns. (b) Geometric difference and union based on the density of the given pattern. (c) Implicit union operation for color editing.}
    \label{fig:ni:boolean}
\end{figure*}
Besides the operations on the explicit part in the tetrahedral neural representation,
inter-operations between different neural radiance fields are essential to certain cases, such as Boolean operations.
To merge or subtract two neural radiance fields, we need to define boolean status on the implicit fields.
Luckily, its much easier to define the boolean status on the implicit field(such as NeRF) than the explicit geometry(such as mesh).
For explicit geometry, the boolean status is defined by the in/out status of the sampling point,
which is not well defined for non-watertight geometry, and usually hard to compute.
On the other hand, the implicit field is indexed with positional parameters in a universal encoding space $G$,
which allows us to define the boolean status based on the scalar values interpreted for each sampling point.
Here we use a straightforward filtering algorithm to define the boolean status $\mathcal{B}(\lambda)$ based on simple thresholds:
\begin{equation}
    \label{eq:boolean}
    \mathcal{B}(\lambda) = \begin{cases}
        1, & \text{if } \mathcal{D}_\sigma(\lambda) > \epsilon \\
        0, & \text{otherwise}
    \end{cases}
\end{equation}
Therefore, the boolean field $\mathcal{B}$ is a binary function of the density field $\mathcal{D}$.
Based on the boolean field, we can define the basic boolean operations between neural radiance fields $\mathcal{B}_i$ and $\mathcal{B}_j$ as follows:
\begin{equation}
    \begin{aligned}
         & \mathcal{B}_i \cup \mathcal{B}_j = \mathcal{B}_i + \mathcal{B}_j - \mathcal{B}_i \cdot \mathcal{B}_j \\
         & \mathcal{B}_i \cap \mathcal{B}_j = \mathcal{B}_i \cdot \mathcal{B}_j                                 \\
         & \mathcal{B}_i \setminus \mathcal{B}_j = \mathcal{B}_i - \mathcal{B}_i \cdot \mathcal{B}_j            \\
         & \mathcal{B}_i \oplus \mathcal{B}_j = \mathcal{B}_i + \mathcal{B}_j (mod 2)
    \end{aligned}
\end{equation}
Then we mask out the corresponding region of the objects and do the local re-training by using the result boolean field as binary weights for the volumetric rendering process.
\vspace{-1ex}
\begin{equation}
    \begin{aligned}
         & C_{k} = \mathsf{D}_r(
        \underbrace{\int_{t_0}^{t_1}T_{tet}(\lambda_t)\underbrace{\mathcal{B}(\lambda_t)}_{\textbf{\textit{boolean}}}\underbrace{\mathsf{D}_{\sigma}(\mathsf{E}_p(\lambda_t))}_{\textit{density}}\cdot \underbrace{\mathsf{E}_p(\lambda_t)}_{\textit{pos feature}} dt}_{\textit{accumlated feature}}
        \oplus\underbrace{\mathsf{E}_d(\vec{d})}_{\textit{dir feature}})
    \end{aligned}
\end{equation}
\vspace{-0.1ex}
As shown in \figref{ni:boolean}, through implicit boolean operations, we can add constructive pattern details to a pre-trained \textit{Neural Impostor} model.
Please note that the boolean field determines the presence of the density field, but the color of the model is interpreted based on the selection of the radiance field.
By the way, boolean operations can be accelerated by first checking the boolean state of tetrahedron and then performing implicit boolean operations on the selected tetrahedron.
By default, boolean operations are binary selective operations on implicit fields.
As a natural extension, selective blending can be achieved using the similar strategy.
Instead of keep or drop a radiance field, we can blend or perform any mutual operations on two radiance fields.
For example, as shown in \figref{ni:boolean} (c), within the region defined by the pattern, we multiply the color given by the pattern with the density field representing fur in the original \textit{Neural Impostor} resulting in a fur model with color. For such cases, local retraining is accomplished by using the second-stage blending perspective-related appearance as the target after determining the tetrahedron covered by the pattern:
\begin{equation}\tag{3-16}
    \begin{aligned}
        C_{k} = & \mathsf{D}_r^i(
        \int_{t_0}^{t_1}T_{tet}^i(\lambda_t^i)\mathcal{B^i}(\lambda_t^i)\mathsf{D}_{\sigma}^i(\mathsf{E}_p^i(\lambda_t^i))\cdot \mathsf{E}_p^i(\lambda_t^i) dt
        \oplus\mathsf{E}_d^i(\vec{d^i})) \\
        +       & \mathsf{D}_r^j(
        \int_{t_0}^{t_1}T_{tet}^j(\lambda_t^j)\mathcal{B^j}(\lambda_t^j)\mathsf{D}_{\sigma}^j(\mathsf{E}_p^j(\lambda_t^j))\cdot \mathsf{E}_p^j(\lambda_t^j) dt
        \oplus\mathsf{E}_d^j(\vec{d^j}))
    \end{aligned}
\end{equation}
\section{Experiments}
\label{sec:expr}
In this chapter, we first present the data acquisition method and implementation details of Neural Impostors.
Then, we perform qualitative and quantitative analyses on the modeling capability and rendering efficiency of \textit{Neural Impostor},  using \textit{nerfstudio} \cite{nerfstudio} as a benchmark for comparison.
% In contrast to the triangle-based modeling approaches employed in \cite{Chen2022MobileNeRFET, Yang2022NeuMeshLD}, the tetrahedral mesh-based modeling approach utilized by \textit{Neural Impostor} offers distinct advantages when dealing with instances that possess intricate volumetric appearances.
To demonstrate this, we create a dataset specifically for plush toys through rendering and real captures and merge it with the \textit{nerf-synthetic} dataset for quantitative analysis.

\subsection{Implementation Details}
\paragraph*{Data Acquisition}
The input for \textit{Neural Impostor} consists of a set of images with corresponding camera parameters and a tetrahedral mesh of the input scene.
The input images are similar to those used in traditional 3D reconstruction tasks and serve as input for the Neural Impostors.
Since the modeling approach of \textit{Neural Impostor} does not require fine geometric details, we have robust support for generating tetrahedral meshes from both generated models in modeling software and reconstructed meshes from scene captures.
Specifically, if the 3D scene is rendered from a virtual model, we can generate a coarse triangular mesh envelope of the original scene by simplifying and decimating the original 3D model.
If the 3D scene is captured from real cameras, we run structure-from-motion algorithms to obtain camera poses and leverage efficient reconstruction methods from "NSR" \cite{instant_nsr2022} to create a rough surface mesh of the scene for the NeRF model.
Then, we compute the triangular mesh envelope based on the surface mesh by using shape modifiers in geometry processing libraries or modeling softwares, such as Libigl and Blender, etc.
This process enables us to generate a concise representation of the surface geometry that encapsulates the volumetric characteristics of the object.
Finally, we use the \textit{TetWild}~\cite{hu2020fast} algorithm to generate the tetrahedral mesh from the triangular mesh envelope.

\paragraph*{Model Structure}
As described in \secref{ni:hashing}, the trainable components in \textit{Neural Impostor} include the multi-tetrahedron hash encoder $\mathsf{E}p$ and the density decoder $\mathsf{D}{\sigma}$ and radiance decoder $\mathsf{D}_{r}$ based on multi-layer perceptrons.
In most scenarios, we fix the overall size of the hash encoding table to $2^{19}$, and the hash table size per tetrahedron varies between $2^8$ and $2^{11}$, depending on the number of tetrahedra.
More specifically, inspired by the approach in "Instant-NGP" \cite{Mller2022InstantNG} for handling different resolution levels, we optimize the hash encoding tables of all tetrahedra in each \textit{Neural Impostor} by packing them together based on tetrahedron indices.
During querying, we locate the fixed region of the packed hash table using the tetrahedron index.
Our density decoder and appearance decoder both adopt a multi-layer perceptron structure.
The density decoder includes a hidden layer with a width of 16, while the appearance decoder includes two hidden layers each with a width of 64.
% an MLP with one hidden layer and a width of 16 as the density decoder and two MLPs with two hidden layers and a width of 64 each as the radiance decoder.
\begin{table}[htp!]\centering
    \caption{Comparison of Rendering Efficiency (conducted on the \textbf{mic} scene in the \textit{nerf-synthetic} dataset, consisting of 294 tetrahedrons and 797 triangles). While NeRF-Editing\cite{nerf-editing2022} takes approximately 7 seconds to render an image with a resolution of $800\times 800$, \textit{Neural Impostor} can achieve real-time rendering at $39.42$FPS with Optix acceleration. Furthermore, it satisfies real-time gaming requirements after baking features onto the surfaces of tetrahedrons and rendering with a rasterizer.}\label{tab:ni:query_speed}
    \scriptsize
    \resizebox{\linewidth}{!}{%
    \begin{tabular}{c|ccc}\toprule
    \textbf{Method} &\textbf{NeRF-Editing} &\textbf{Optix Accelerated} &\textbf{Baking+Rasterization} \\\hline
    \textbf{FPS }$\uparrow$ &0.147 &39.42 &\cellcolor[HTML]{f1f0fc}\textbf{157.83} \\
    \bottomrule
    \end{tabular}
    }
    \end{table}

\begin{figure*}[t]
    \centering
    \def\svgwidth{2\columnwidth}
    %% Creator: Inkscape 1.2.2 (b0a8486541, 2022-12-01), www.inkscape.org
%% Accompanies image file 'quality_eval_furry.pdf' (pdf, eps, ps)
\begingroup%
  \makeatletter%
  \providecommand\color[2][]{%
    \errmessage{(Inkscape) Color is used for the text in Inkscape, but the package 'color.sty' is not loaded}%
    \renewcommand\color[2][]{}%
  }%
  \providecommand\transparent[1]{%
    \errmessage{(Inkscape) Transparency is used (non-zero) for the text in Inkscape, but the package 'transparent.sty' is not loaded}%
    \renewcommand\transparent[1]{}%
  }%
  \providecommand\rotatebox[2]{#2}%
  \newcommand*\fsize{\dimexpr\f@size pt\relax}%
  \newcommand*\lineheight[1]{\fontsize{\fsize}{#1\fsize}\selectfont}%
  \ifx\svgwidth\undefined%
    \setlength{\unitlength}{876bp}%
    \ifx\svgscale\undefined%
      \relax%
    \else%
      \setlength{\unitlength}{\unitlength * \real{\svgscale}}%
    \fi%
  \else%
    \setlength{\unitlength}{\svgwidth}%
  \fi%
  \global\let\svgwidth\undefined%
  \global\let\svgscale\undefined%
  \makeatother%
  \begin{picture}(1,0.79223744)%
    \lineheight{1}%
    \setlength\tabcolsep{0pt}%
    \put(0,0){\includegraphics[width=\unitlength,page=1]{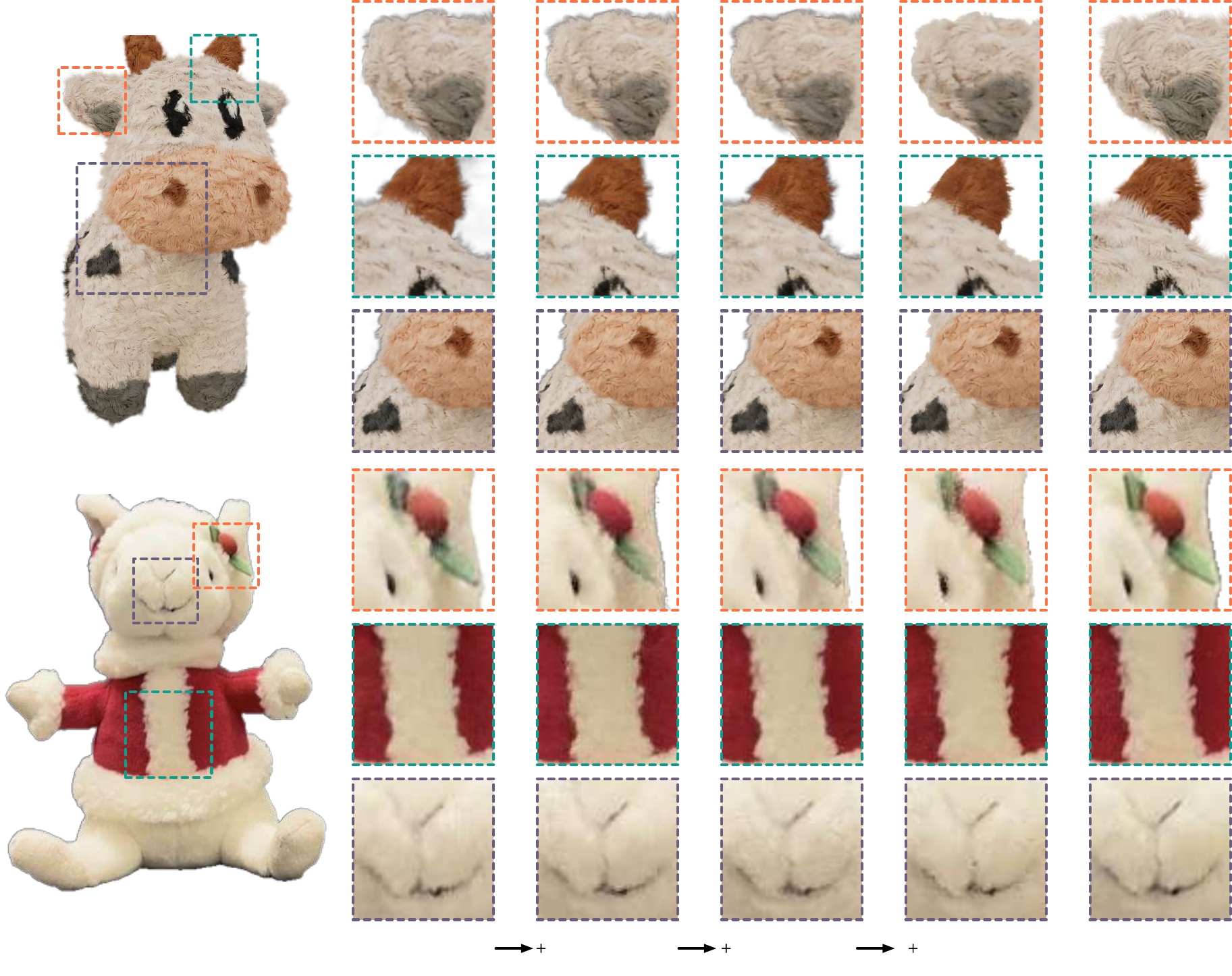}}%
    \put(0.13706524,0.41616041){\color[rgb]{0,0,0}\makebox(0,0)[t]{\lineheight{1.25}\smash{\begin{tabular}[t]{c}\large{\textbf{Spot plush}}\end{tabular}}}}%
    \put(0.14214872,0.03432008){\color[rgb]{0,0,0}\makebox(0,0)[t]{\lineheight{1.25}\smash{\begin{tabular}[t]{c}\large{\textbf{Lamb plush}}\end{tabular}}}}%
    \put(0.34360731,0.02545284){\color[rgb]{0,0,0}\makebox(0,0)[t]{\lineheight{1.25}\smash{\begin{tabular}[t]{c}Nerfstudio\\(instant-ngp bounded)\end{tabular}}}}%
    \put(0.4929124,0.02545284){\color[rgb]{0,0,0}\makebox(0,0)[t]{\lineheight{1.25}\smash{\begin{tabular}[t]{c}Barycentric\\Encoding\end{tabular}}}}%
    \put(0.64245578,0.02545284){\color[rgb]{0,0,0}\makebox(0,0)[t]{\lineheight{1.25}\smash{\begin{tabular}[t]{c}Barycentric\\Sampling\end{tabular}}}}%
    \put(0.79212039,0.02560961){\color[rgb]{0,0,0}\makebox(0,0)[t]{\lineheight{1.25}\smash{\begin{tabular}[t]{c}Feature\\Integration\end{tabular}}}}%
    \put(0.94184144,0.0246063){\color[rgb]{0,0,0}\makebox(0,0)[t]{\lineheight{1.25}\smash{\begin{tabular}[t]{c}Ground\\Truth\end{tabular}}}}%
  \end{picture}%
\endgroup%

    \caption{Visual comparison on the plush toy dataset.
    The \textit{spot} instance above is synthetic data, while the \textit{lamb} instance below is captured from the real-world with a camera.
    \textit{Neural Impostor} accurately reconstructs the hair details.
    On the instance of the \textit{lamb}, barycentric encoding achieves better quality on recovering the highly saturated parts (the red part) of the scene than regular space encoding. At the same time, on the instance of the \textit{spot}, tetrahedral meshes help restrict the sampling space and thus better recover the spatial density field, eliminating the \textit{black edge} phenomenon.}
    \label{fig:ni:vis_comp_furry}
\end{figure*}
\smallskip
\begin{figure*}[t]
    \centering
    \def\svgwidth{2\columnwidth}
    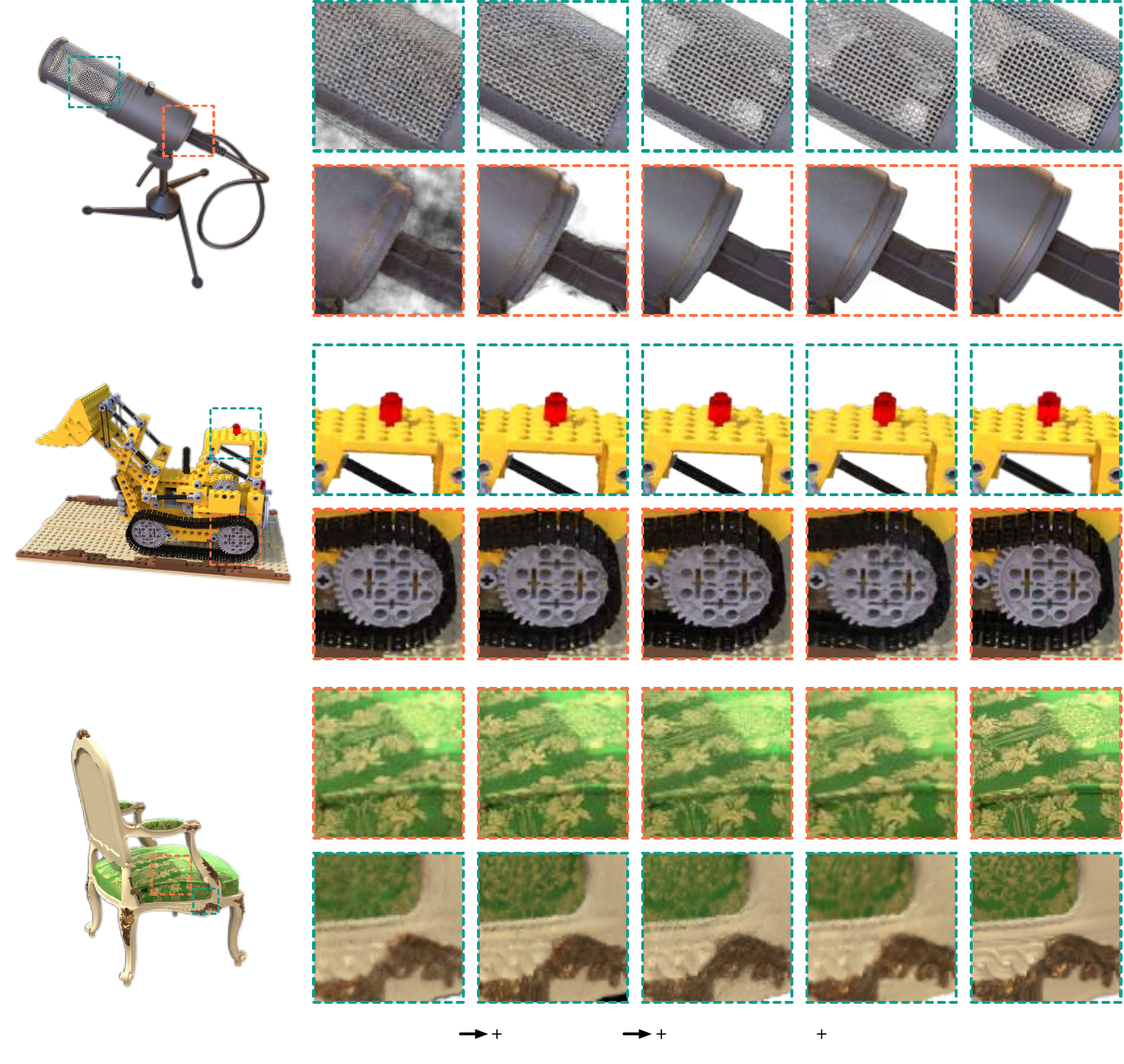

    \caption{Visual comparison on the \textit{nerf-synthetic} dataset.
    \textit{Neural Impostor} provide comparable or even better quality.
    More specifically, \textit{Neural Impostor} performs better on disentangling the density field and the radiance field.
    For example, in the \textit{mic} instance, with the help of barycentric sampling, we recover not only the details of the metal case but also the capsule of the microphone.}
    \label{fig:ni:vis_comp_nerf}
\end{figure*}

% \subsection{Quantitative Evaluation}
\subsection{Reconstruction Quality Evaluation}
% \subsubsection{\rev{Reconstruction Quality}}
% \cmt{Rewrite, add details for each case.}
% \subsubsection{\rev{Cage Quality}}
% % \cmt{Cube, BBox, Ico_sub0-4, Original}

% \subsubsection{\rev{Animation Quality}}
% \cmt{Merge from previous.}

\begin{table*}[tb!]
    \renewcommand*{\arraystretch}{1.2}
    \centering
    \caption{Integrated testing result based on nerfstudio}\label{tab:ni:quality}
    \scriptsize
    \resizebox{\linewidth}{!}{%
    \begin{tabular}{c|cc|cc|cc|cc|cc}\toprule
        \multirow{2}{*}{\textbf{Method}} &\multicolumn{2}{c|}{} &\multicolumn{4}{c|}{\textbf{Nerf-Synthetic (Avg. Impostor Size = 545.57)}} &\multicolumn{4}{c}{\textbf{Plush Toy (Average Impostor Size = 1754.4)}} \\
        &BatchSize  &Steps &Training Time   &Rendering FPS  &PSNR~$\uparrow$  &SSIM~$\uparrow$   &Training Time  &Rendering FPS  &PSNR~$\uparrow$   &SSIM~$\uparrow$ \\\hline
        NGP-Bounded &8192 &30k             &40.09m         &33.272 &\cellcolor[HTML]{f1f0fc}\textbf{36.545} &0.981 &44.65m &22.28 &32.708 &0.9474 \\
        BaryEnc     &8192 &30k             &36.72m         &33.134 &36.473 &0.9809 &32.51m &26.95 &32.843 &0.947 \\
        BarySpl     &8192 &30k             &30.63m         &33.566 &36.423 &\cellcolor[HTML]{f1f0fc}\textbf{0.9813} &29.91m &27.51 &\cellcolor[HTML]{f1f0fc}\textbf{32.915} &\cellcolor[HTML]{f1f0fc}\textbf{0.9477} \\
        FeatInt &8192 &30k &\cellcolor[HTML]{f1f0fc}\textbf{29.17m} &\cellcolor[HTML]{efefef}39.42 &34.53 &0.9764 &\cellcolor[HTML]{f1f0fc}\textbf{28.74m} &\cellcolor[HTML]{efefef}33.11 &32.797 &0.9393 \\
        Baking + Rast &/ &/ &/ &\cellcolor[HTML]{f1f0fc}\textbf{157.83} &33.796 &0.9695 &/ &\cellcolor[HTML]{f1f0fc}\textbf{129.32} &32.249 &0.937 \\
        \bottomrule
        \end{tabular}
    }
\end{table*}
As mentioned earlier, we analyze the modeling capability of \textit{Neural Impostor} from the perspectives of modeling, rendering efficiency, and continuity in the deformation process.
In this section, we focus on evaluating the reconstruction quality.
Specifically, using the implementation of \textit{Instant-NGP} in \textit{nerfstudio}\cite{nerfstudio} (method "NGP-Bounded" in \tabref{ni:quality}) as a baseline, we first replace the Cartesian coordinate-based encoding part in the original model with barycentric coordinate encoding while retaining the sampling method based on occupancy fields (method "BaryEnc" in \tabref{ni:quality}).
Then, we replace the sampling method based on occupancy fields with barycentric coordinate-based sampling (method "BarySpl" in \tabref{ni:quality}).
Finally, we integrate the "early integration" rendering approach from \secref{ni:rendering} into the model based on the barycentric coordinate sampling method (method "FeatInt" in \tabref{ni:quality}).
We also provide the results of rendering after baking onto the tetrahedral surface and using rasterization for rendering (method \textit{Baking+Rast} in \tabref{ni:quality}).
Among them, we compare the PSNR (Peak Signal-to-Noise Ratio) and SSIM (Structural Similarity) under various rendering methods for 7 instances in the \textit{nerf-synthetic} dataset and 5 instances in the \textit{Plush Toy} dataset (including 3 real-captured instances and 2 rendered instances).
PSNR measures the quality of an image by computing the mean squared error between the original image and the processed (compressed or reconstructed) image.
The higher the PSNR value, the smaller the difference between the processed image and the original image, indicating better quality.
SSIM measures the similarity between two images by considering their luminance, contrast, and structural information.
It is a more accurate reflection of image quality compared to traditional metrics like PSNR, as it aligns better with human visual perception.
We provide visual comparisons of the reconstruction results on the \textit{nerf-synthetic} and plush toy datasets in
% \figref{ni:vis_comp_nerf} and \figref{ni:vis_comp_furry}, respectively.
\figref{ni:vis_comp_nerf} and \figref{ni:vis_comp_furry},
as well as their quantitative results in \tabref{ni:quality}.
The quantitative analysis on the \textit{nerf-synthetic} dataset shows that, without introducing the \textit{early integration} rendering approach (\secref{ni:rendering}), using the barycentric coordinate encoding in the modeling process (method "BaryEnc") achieves comparable quality to the original \textit{Instant-NGP} \cite{Mller2022InstantNG}, with a PSNR value of 33.134 for barycentric coordinate encoding compared to 33.272 for the original \textit{Instant-NGP} encoding. Furthermore, since barycentric coordinate encoding is not limited by resolution, it can achieve good reconstruction quality (PSNR = $32.915$) in complex scenes such as plush toys when combined with barycentric coordinate-based sampling, surpassing the quality of the original \textit{Instant-NGP} (PSNR = $32.708$).
When the \textit{early integration} (\secref{ni:rendering}) approach is introduced, although rendering efficiency improves, there is a slight decrease in modeling quality due to the texture features of each tetrahedron being baked onto the tetrahedral surface. However, after baking, we can use rasterization for rendering, resulting in a significant improvement in rendering efficiency (from $30+$ FPS before baking to $100+$ FPS after baking). Therefore, in scenarios with high interactivity demands, we can choose to use the \textit{early integration} (\secref{ni:rendering}) rendering approach to improve interactive performance.
\begin{figure*}[t]
    \centering
    \def\svgwidth{2\columnwidth}
    \import{./figs/}{tet_quality.pdf_tex}

    \caption{The impact of tetrahedron proxy quality. The fluffy sphere is reconstructed with different tetrahedron proxies. The visual quality and PSNR both have subtle differences, indicating the insignificant impact of the proxy quality.}
    \label{fig:ni:cage_quality}
\end{figure*}

The proposed hybrid representation relies on an explicit tetrahedron mesh as a proxy of the neural implicit field.
The quality and density of tetrahedron mesh do impact the encoding of the neural radiance field.
However, these factors only have subtle effects on the reconstruction quality.
% Basically, we utilize the tetrahedron mesh for efficient spatial encoding, and more importantly, we use the tetrahedron mesh as an explicit proxy for the different editing operations.
Figure~\ref{fig:ni:cage_quality} examines the impact of different tetrahedral mesh quality and density on the reconstruction quality of neural impostors in static scene reconstruction.
We tests 7 different tetrahedron proxies with similar implicit field sizes (i.e. sum of hash table size of all tetrahedrons) on the fluffy ball example by varying the tetrahedron density and envelope shape.
The similar PSNR of all tests indicates that the selection of tetrahedron proxies has an insignificant influence on the quality of the implicit field.
\subsection{Editing Quality Evaluation}
Different from evaluating the reconstruction quality, there is no well-defined metric or reference ground truth for the edited radiance field.
Therefore, we use the rendering of the edited Neural Impostor as a qualitative evaluation(please check \secref{phys_sim}) and calculate the LPIPS (Learned Perceptual Image Patch Similarity) between the edited Neural Impostor to the reference rendering as the quantitative evaluation.
Here, we analyzed the rendering quality during simulation. Specifically, we use the Houdini simulation engine to create dynamic meshes for fracture and deformation effects for the testing scenes. We then perform animation simulations using the ray bending algorithm in NeRF-Editing\cite{nerf-editing2022} with two-stage sampling (\textit{Ray Bending} in \tabref{ni:anim_quality}), the algorithm based on occupancy field sampling combined with barycentric coordinate encoding (\textit{Occupancy Field} in \tabref{ni:anim_quality}), and the barycentric-based sampling algorithm in \secref{ni:sampling} (\textit{Barycentric Sampling} in \tabref{ni:anim_quality}). Since we cannot obtain real multi-view data as a reference during the simulation process, and the deformations generated during the simulation can be negligible with a sufficiently high frame rate, we calculate the LPIPS between each pair of frames as a measure of animation simulation quality.
% It is worth to note that NeRF-Editing\cite{nerf-editing2022}, most instances only have surface appearance, so the ray bending method does not involve sampling in the undefined regions outside the surface mesh during the actual sampling process. 
% However, for objects with complex volumetric appearances, the actual tetrahedral grid is often slightly larger than the real surface of the object. The starting points in the sampling process are actually defined by the boundaries of the tetrahedral grid, which introduces points with poorly defined densities in space. As a result, black borders may appear at the mesh boundaries, affecting the PSNR even in static situations (please refer to the supplemental material video for details). 
From the comparison of LPIPS between each pair of frames in \tabref{ni:anim_quality}, it can be observed that our barycentric sampling method better maintains the appearance stability during the deformation process.
\begin{table}[htp!]
\label{tbl:ni:anim_comp}
\centering
    \caption{Animation quality comparison. We calculate the LPIPS value between adjancent sampled frames during deformation, and observed that }\label{tab:ni:anim_quality}
    \scriptsize
    \resizebox{\linewidth}{!}{%
    \begin{tabular}{c|p{0.1\columnwidth}|p{0.1\columnwidth}|p{0.1\columnwidth}}\toprule
    \textbf{Method} &\textbf{Deform LPIPS} &\textbf{Reference Frame PSNR} &\textbf{Reference Frame SSIM} \\\hline
    \textbf{Ray Bending}    &0.0999 &27.87 &0.7572 \\
    \textbf{Occupancy Field} &0.0876 &\cellcolor[HTML]{f1f0fc}\textbf{37.01} &\cellcolor[HTML]{f1f0fc}\textbf{0.9233} \\
    \textbf{Barycentric Sampling (Ours)} &\cellcolor[HTML]{f1f0fc}\textbf{0.0669} &36.97 &0.9175 \\
    \bottomrule
    \end{tabular}
    }
\end{table}

\section{Application}
\label{sec:application}

\begin{figure*}
    \centering
    \def\svgwidth{2\columnwidth}
    %% Creator: Inkscape 1.2.2 (b0a8486541, 2022-12-01), www.inkscape.org
%% Accompanies image file 'vis_deform_1.pdf' (pdf, eps, ps)
\begingroup%
  \makeatletter%
  \providecommand\color[2][]{%
    \errmessage{(Inkscape) Color is used for the text in Inkscape, but the package 'color.sty' is not loaded}%
    \renewcommand\color[2][]{}%
  }%
  \providecommand\transparent[1]{%
    \errmessage{(Inkscape) Transparency is used (non-zero) for the text in Inkscape, but the package 'transparent.sty' is not loaded}%
    \renewcommand\transparent[1]{}%
  }%
  \providecommand\rotatebox[2]{#2}%
  \newcommand*\fsize{\dimexpr\f@size pt\relax}%
  \newcommand*\lineheight[1]{\fontsize{\fsize}{#1\fsize}\selectfont}%
  \ifx\svgwidth\undefined%
    \setlength{\unitlength}{5734.00012207bp}%
    \ifx\svgscale\undefined%
      \relax%
    \else%
      \setlength{\unitlength}{\unitlength * \real{\svgscale}}%
    \fi%
  \else%
    \setlength{\unitlength}{\svgwidth}%
  \fi%
  \global\let\svgwidth\undefined%
  \global\let\svgscale\undefined%
  \makeatother%
  \begin{picture}(1,0.61580048)%
    \lineheight{1}%
    \setlength\tabcolsep{0pt}%
    \put(0,0){\includegraphics[width=\unitlength,page=1]{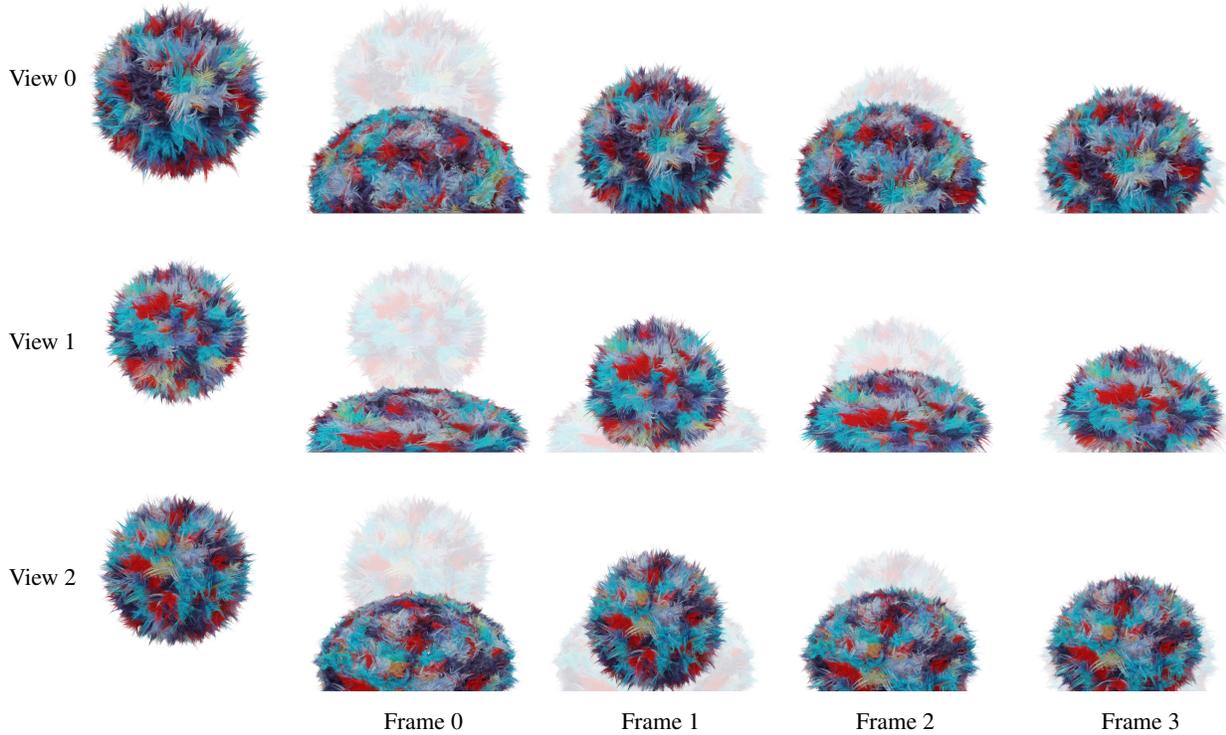}}%
    \put(0.01964549,0.52772969){\color[rgb]{0,0,0}\makebox(0,0)[lt]{\lineheight{1.25}\smash{\begin{tabular}[t]{l}View 0\end{tabular}}}}%
    \put(0.01970936,0.31991){\color[rgb]{0,0,0}\makebox(0,0)[lt]{\lineheight{1.25}\smash{\begin{tabular}[t]{l}View 1\end{tabular}}}}%
    \put(0.01971894,0.13432441){\color[rgb]{0,0,0}\makebox(0,0)[lt]{\lineheight{1.25}\smash{\begin{tabular}[t]{l}View 2\end{tabular}}}}%
    \put(0.31537417,0.02085235){\color[rgb]{0,0,0}\makebox(0,0)[lt]{\lineheight{1.25}\smash{\begin{tabular}[t]{l}Frame 0\end{tabular}}}}%
    \put(0.50232119,0.02091941){\color[rgb]{0,0,0}\makebox(0,0)[lt]{\lineheight{1.25}\smash{\begin{tabular}[t]{l}Frame 1\end{tabular}}}}%
    \put(0.68726796,0.02085235){\color[rgb]{0,0,0}\makebox(0,0)[lt]{\lineheight{1.25}\smash{\begin{tabular}[t]{l}Frame 2\end{tabular}}}}%
    \put(0.88060208,0.02085235){\color[rgb]{0,0,0}\makebox(0,0)[lt]{\lineheight{1.25}\smash{\begin{tabular}[t]{l}Frame 3\end{tabular}}}}%
  \end{picture}%
\endgroup%

    \caption{Real-time Soft Body Deformation Utilizing \textit{Neural Impostor}: The \textit{Neural Impostor} representation enables us to sustain consistent, high-quality rendering amidst deformation. This is applicable even for objects exhibiting intricate volumetric appearances.}
    \label{fig:ni:vis_deform}
\end{figure*}
\begin{figure*}
    \centering
    \def\svgwidth{2\columnwidth}
    %% Creator: Inkscape 1.2.2 (732a01da63, 2022-12-09), www.inkscape.org
%% Accompanies image file 'vis_deform_2.pdf' (pdf, eps, ps)
\begingroup%
  \makeatletter%
  \providecommand\color[2][]{%
    \errmessage{(Inkscape) Color is used for the text in Inkscape, but the package 'color.sty' is not loaded}%
    \renewcommand\color[2][]{}%
  }%
  \providecommand\transparent[1]{%
    \errmessage{(Inkscape) Transparency is used (non-zero) for the text in Inkscape, but the package 'transparent.sty' is not loaded}%
    \renewcommand\transparent[1]{}%
  }%
  \providecommand\rotatebox[2]{#2}%
  \newcommand*\fsize{\dimexpr\f@size pt\relax}%
  \newcommand*\lineheight[1]{\fontsize{\fsize}{#1\fsize}\selectfont}%
  \ifx\svgwidth\undefined%
    \setlength{\unitlength}{5588bp}%
    \ifx\svgscale\undefined%
      \relax%
    \else%
      \setlength{\unitlength}{\unitlength * \real{\svgscale}}%
    \fi%
  \else%
    \setlength{\unitlength}{\svgwidth}%
  \fi%
  \global\let\svgwidth\undefined%
  \global\let\svgscale\undefined%
  \makeatother%
  \begin{picture}(1,0.61724771)%
    \lineheight{1}%
    \setlength\tabcolsep{0pt}%
    \put(0,0){\includegraphics[width=\unitlength,page=1]{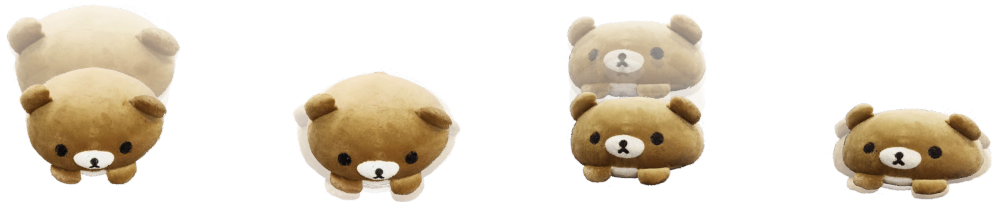}}%
    \put(0.00463883,0.29186667){\color[rgb]{0,0,0}\makebox(0,0)[lt]{\lineheight{1.25}\smash{\begin{tabular}[t]{l}\textbf{Spot Plush}\end{tabular}}}}%
    \put(0.00406212,0.25886982){\color[rgb]{0,0,0}\makebox(0,0)[lt]{\lineheight{1.25}\smash{\begin{tabular}[t]{l}\textbf{Bear Plush}\end{tabular}}}}%
    \put(0.09520306,0.29186667){\color[rgb]{0.99215686,0.44313725,0.27058824}\makebox(0,0)[lt]{\lineheight{1.25}\smash{\begin{tabular}[t]{l}(Synthetic data)\end{tabular}}}}%
    \put(0.09520306,0.25886982){\color[rgb]{0.04313725,0.61568627,0.55294118}\makebox(0,0)[lt]{\lineheight{1.25}\smash{\begin{tabular}[t]{l}(Captured data)\end{tabular}}}}%
  \end{picture}%
\endgroup%

    \caption{\textit{Neural Impostor} in Action - Real-Time Deformation of Plush Toys: Our \textit{Neural Impostor} technique excels at performing seamless deformation for plush toys, which are notoriously difficult to model using traditional triangle mesh representation due to their fluffy appearances. This advancement enables high-quality, consistent rendering regardless of the object's complexity..}
    \label{fig:ni:vis_deform_cont}
\end{figure*}

\begin{figure*}
    \centering
    \def\svgwidth{1.7\columnwidth}
    \import{./figs/}{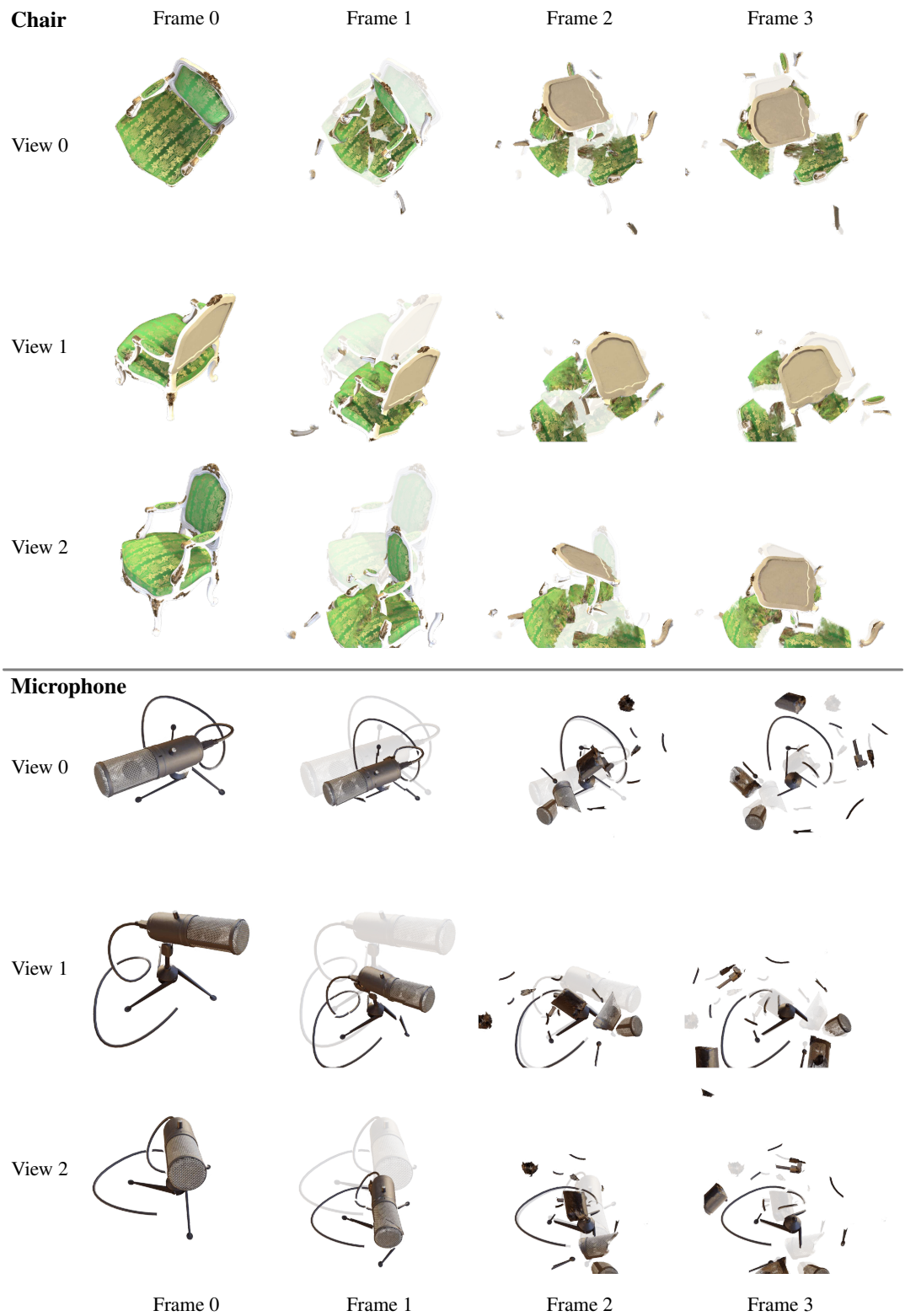}

    \caption{Plastic fracture with \textit{Neural Impostor}.}
    \label{fig:ni:vis_frag}
\end{figure*}

\begin{figure*}
    \centering
    \def\svgwidth{2\columnwidth}
    %% Creator: Inkscape 1.2.2 (732a01da63, 2022-12-09), www.inkscape.org
%% Accompanies image file 'snowman.pdf' (pdf, eps, ps)
\begingroup%
  \makeatletter%
  \providecommand\color[2][]{%
    \errmessage{(Inkscape) Color is used for the text in Inkscape, but the package 'color.sty' is not loaded}%
    \renewcommand\color[2][]{}%
  }%
  \providecommand\transparent[1]{%
    \errmessage{(Inkscape) Transparency is used (non-zero) for the text in Inkscape, but the package 'transparent.sty' is not loaded}%
    \renewcommand\transparent[1]{}%
  }%
  \providecommand\rotatebox[2]{#2}%
  \newcommand*\fsize{\dimexpr\f@size pt\relax}%
  \newcommand*\lineheight[1]{\fontsize{\fsize}{#1\fsize}\selectfont}%
  \ifx\svgwidth\undefined%
    \setlength{\unitlength}{1684bp}%
    \ifx\svgscale\undefined%
      \relax%
    \else%
      \setlength{\unitlength}{\unitlength * \real{\svgscale}}%
    \fi%
  \else%
    \setlength{\unitlength}{\svgwidth}%
  \fi%
  \global\let\svgwidth\undefined%
  \global\let\svgscale\undefined%
  \makeatother%
  \begin{picture}(1,0.5391924)%
    \lineheight{1}%
    \setlength\tabcolsep{0pt}%
    \put(0,0){\includegraphics[width=\unitlength,page=1]{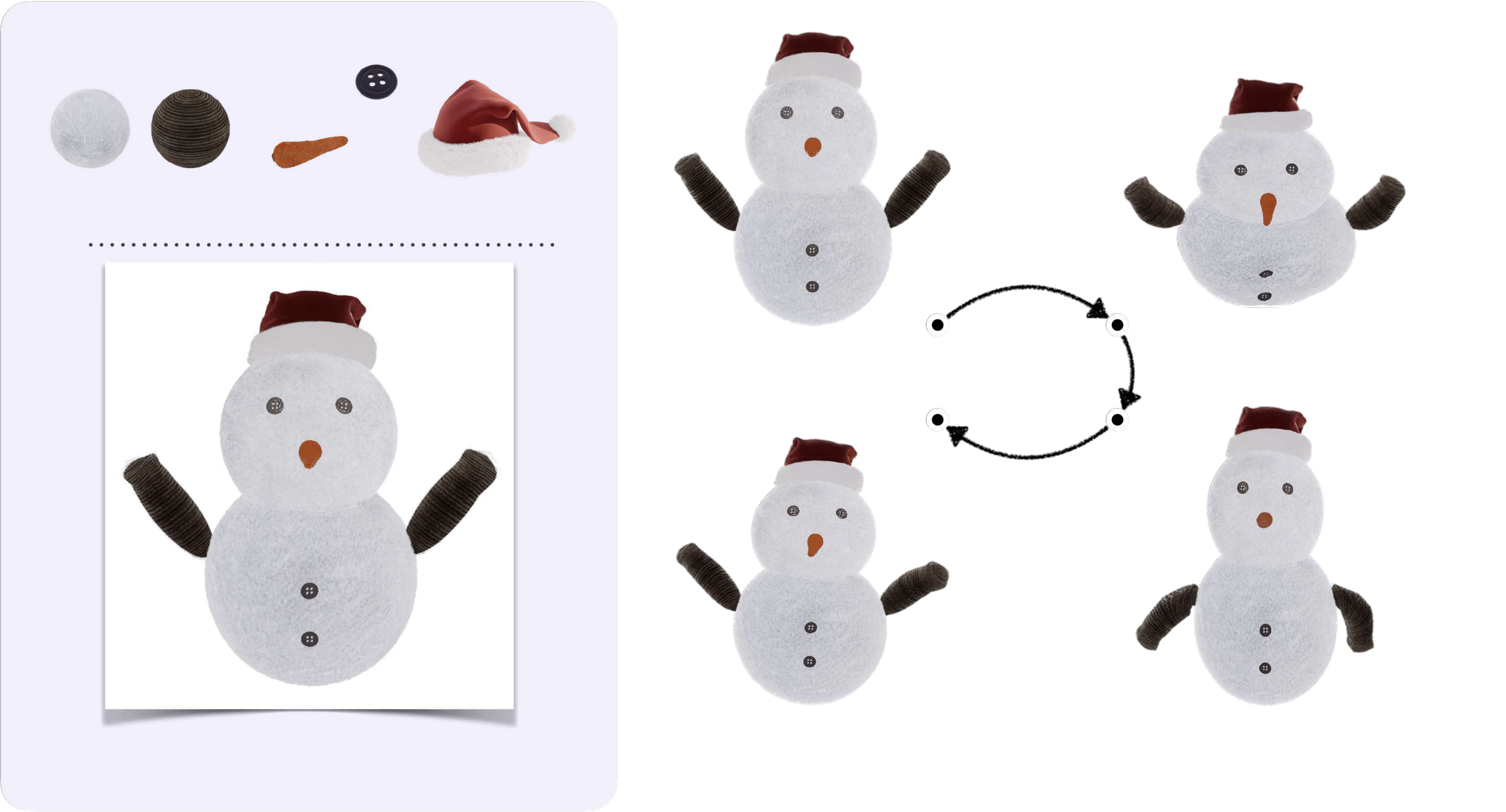}}%
    \put(0.19907501,0.38932375){\color[rgb]{0,0,0}\makebox(0,0)[t]{\lineheight{1.25}\smash{\begin{tabular}[t]{c}(a) Pre-trained neural impostors\end{tabular}}}}%
    \put(0.19974915,0.02741729){\color[rgb]{0,0,0}\makebox(0,0)[t]{\lineheight{1.25}\smash{\begin{tabular}[t]{c}(b) Composed \textbf{Snowman}\end{tabular}}}}%
    \put(0.69650276,0.02749703){\color[rgb]{0,0,0}\makebox(0,0)[t]{\lineheight{1.25}\smash{\begin{tabular}[t]{c}(c) Animating \textbf{Snowman}\end{tabular}}}}%
  \end{picture}%
\endgroup%

    \caption{Authoring and simulation based on Neural Impostors.
    (a) Pre-trained \textit{Neural Impostor}s as building blocks of a (b)\textbf{snowman}.
    (c) Animated frames of the \textbf{snowman} model by rendering the composed \textit{Neural Impostor}.
    % We simulated the creative process in traditional 3D modeling software using material balls and pre-modeled Neural Impostors as material. By retraining, we can apply the appearance represented by the material ball to a simple tetrahedral mesh to build the body of the snowman. Then, by deforming and simulating the Neural Impostor that applies the material ball appearance, we created the arms of the snowman. Finally, by combining with the pre-trained Neural Impostor, we created the final snowman model. It is worth mentioning that the result of the above operation is still a Neural Impostor model, which allows us to apply all editing or deformation operations in the Neural Impostor.
    }
    \label{fig:ni:snowman}
\end{figure*}

\subsection{Physical Simulation}
\label{sec:phys_sim}
Soft body deformation algorithms are used in computer graphics to simulate the motion and deformation of soft, flexible materials such as cloth, rubber, and muscles.
These algorithms combine physics simulation with numerical methods to calculate the motion and deformation of soft bodies under external forces.
The basic idea of soft body deformation is to discretize the object into many small elements, and solve for the displacement for each node in the elements.
The motion and deformation of each element are calculated based on the forces acting on it, such as gravity, collision forces, tension, and compression forces.
One common method of soft body deformation is using finite element analysis (FEA), which divides the body into small triangles or tetrahedral elements.
Each element is assigned a set of physical properties, such as stiffness and density, which are used to calculate the motion and deformation of the element under external forces.
Other soft body deformation algorithms include particle-spring systems, which model the body as an interconnected network of springs, and position-based dynamics, which use simplified equations to simulate the motion and deformation of the body.
By using the same discretized tetrahedron mesh, we can achieve physical simulation directly on \textit{Neural Impostor}s.
\figref{ni:vis_deform} shows an example of soft body simulation based on \textit{Neural Impostor}.
In this example, we use a furball with a complex appearance but relatively simple geometry (consisting of 204 vertices and 600 tetrahedra) as the object of simulation.
Three significantly different viewpoints and five dynamic frames are extracted and displayed.
From the figure, it can be observed that the \textit{Neural Impostor} can achieve continuous simulation with large deformations while maintaining the quality of the volumetric appearance.
Furthermore, in \figref{ni:vis_deform_cont}, we demonstrate continuous soft body transformations based on rendered data and real-shot data. \textit{Plush Spot} represents the rendered data, while \textit{Plush Bear} represents the real-shot data. For more examples, please refer to the supplementary video material.

In addition to soft body simulation, plastic fracture simulation is a technique used in computer graphics to simulate the destruction of buildings, vehicles, and other structures.
In plastic fracture, each part of the object is treated as a separate plastic element that can be influenced by external forces such as explosions or collisions.
These plastic elements are simulated using physics-based algorithms to calculate their motion and deformation over time.
One common method for simulating plastic fracture is to use Voronoi splitting to break the object into smaller fragments.
These fragments are then simulated using a physics engine to calculate their motion in the scene and collisions with other objects.
Other techniques used for plastic fracture include finite element analysis (FEA) and dynamic fracture modeling, which can produce more detailed and realistic simulations of object destruction.
\figref{ni:vis_frag} illustrates two examples of plastic fracture simulation on \textit{Neural Impostor}s.
Similar to soft body simulation, we utilize the Houdini engine with the original tetrahedral mesh as a proxy for the plastic fracture simulation.

\subsection{Content Authoring with Neural Impostors}
In addition to its robust support for vertex-based animation, as shown in \figref{ni:boolean}, the \textit{Neural Impostor} algorithm further enables Boolean appearance editing (e.g., baking pattern colors onto an existing \textit{Neural Impostor} model as shown in \figref{ni:boolean}(d)) and geometry editing (e.g., sculpting the original \textit{Neural Impostor} as shown in \figref{ni:boolean}(c)) through local retraining operations described in \secref{operations}.
By using explicit tetrahedral meshes as proxies, we can effortlessly combine multiple \textit{Neural Impostor} models to create new scenes, offering the possibility of using \textit{Neural Impostor} as a novel modeling primitive.
\figref{ni:snowman} illustrates the process of creating a new \textit{Neural Impostor} asset from material spheres and three existing \textit{Neural Impostor} models as shown in \figref{ni:snowman}(a) through appearance mapping (snowman body), composition (hat, nose, eyes, and other decorations), and geometry editing (snowman arms).
Specifically, starting from material spheres representing the appearance of a snowball and wooden arms, we transfer the appearance information represented by their materials to the snowman body and arms, represented by a simple spherical geometry, through local retraining.
Additionally, we elongate the arm vertices along the axis to assemble the snowman body.
By combining the constructed body parts with \textit{Neural Impostor} representing the hat and nose, we create a complete snowman model as shown in \figref{ni:snowman}(b).
It is worth noting that, apart from the retraining step involved in appearance mapping, this modeling process aligns completely with traditional 3D modeling software workflows.
After modeling, the snowman remains a new \textit{Neural Impostor} model, allowing seamless integration with subsequent animation simulations.
\figref{ni:snowman}(c) illustrates the process of soft body simulation applied to the snowman model.

The above example is just one of many possibilities. In practical applications, the \textit{Neural Impostor} technology can be used to create various 3D models, including animals, buildings, vehicles, and more.
% By using pre-built \textit{Neural Impostor}s, we can construct a snowman model.
% Through various deformations and transformations such as stretching, twisting, and scaling, we can make the model more natural and realistic in shape.
% In addition to creating models using individual \textit{Neural Impostor}s, we can also combine multiple \textit{Neural Impostor}s in different ways to build more complex and large-scale scenes, where each \textit{Neural Impostor} in the scene.
Additionally, inspired by the construction of neural radiation fields from multi-view images, artists can start appearance modeling directly from real captured images, simplifying the material design process in traditional modeling.
% This approach greatly improves the efficiency and creativity of 3D modeling, enabling designers to express their ideas more freely.
In a nutshell, the \textit{Neural Impostor} technology provides a brand new workflow for authoring complex models with high quality volumetric appearance.
\section{Conclusion}
In this article, we have presented \textit{Neural Impostor}, a hybrid neural representation.
\textit{Neural Impostor} addresses the challenges of editability in neural radiance fields, providing a novel approach to modeling and editing of NeRF.
Unlike previous methods, \textit{Neural Impostor}  employing an explicit tetrahedral mesh and locally encoded radiance fields to achieve high-quality modeling and rendering.
The barycentric-based sampling method ensures rendering quality and efficiency, while the proposed local retraining method allows for fine-grained simulations and editing.
We also demonstrated its applications in various content authoring processes, including physical simulation, geometric editing and model composition.
Meanwhile, there are some aspects that can be improved.
Currently, \textit{Neural Impostor} only supports geometric editing operations.
Extending the range of compatible operations, including relighting and material properties, would be some challenging open problems.
Overall, \textit{Neural Impostor} introduces new possibilities in neural representation, enhancing the modeling, rendering, and editing capabilities of neural primitives in various applications.
% Furthermore, the potential of \textit{Neural Impostor} as a new material paradigm in game development has been discussed. \textit{Neural Impostor} provide high modeling accuracy and real-time rendering capabilities for complex objects, while also offering flexibility in animation production and physics simulations.
% By seamlessly integrating with existing workflows, \textit{Neural Impostor} empower game developers to create more immersive game scenes and characters.
% While \textit{Neural Impostor} currently has some limitations, ongoing advancements, such as the exploration of generative models and semantic neural networks, are expected to further expand the capabilities of \textit{Neural Impostor}, enabling reconstruction, editing, and generation tasks based on single images or videos.
% Overall, \textit{Neural Impostor} presents a promising avenue for pushing the boundaries of 3D modeling and game development.

%-------------------------------------------------------------------------
% bibtex
\bibliographystyle{eg-alpha-doi} 
\bibliography{references}       

% biblatex with biber
% \printbibliography                

%-------------------------------------------------------------------------
\newpage

% \begin{figure*}[tbp]
%   \centering
%   \mbox{} \hfill
%   % the following command controls the width of the embedded PS file
%   % (relative to the width of the current column)
%   \includegraphics[width=.3\linewidth]{sampleFig}
%   % replacing the above command with the one below will explicitly set
%   % the bounding box of the PS figure to the rectangle (xl,yl),(xh,yh).
%   % It will also prevent LaTeX from reading the PS file to determine
%   % the bounding box (i.e., it will speed up the compilation process)
%   % \includegraphics[width=.3\linewidth, bb=39 696 126 756]{sampleFig}
%   \hfill
%   \includegraphics[width=.3\linewidth]{sampleFig}
%   \hfill \mbox{}
%   \caption{\label{fig:ex3}%
%            For publications with color tables (i.e., publications not offering
%            color throughout the paper) please \textbf{observe}: 
%            for the printed version -- and ONLY for the printed
%            version -- color figures have to be placed in the last page.
%            \newline
%            For the electronic version, which will be converted to PDF before
%            making it available electronically, the color images should be
%            embedded within the document. Optionally, other multimedia
%            material may be attached to the electronic version. }
% \end{figure*}

\end{document}